\documentstyle[psfig]{mn}

\title[Physical parameters of faint galaxies] {The physical parameters of 
the evolving population of faint galaxies}

\author[K. Glazebrook \etal]  {Karl Glazebrook,$^1$ Roberto
Abraham,$^2$ Basilio Santiago,$^{2,3}$ Richard Ellis,$^2$ \newauthor 
and Richard Griffiths$^4$\\
$^1$ Anglo-Australian Observatory, PO Box 296, Epping, NSW 2121, AUSTRALIA\\
$^2$ Institute of Astronomy, Madingley Road,
Cambridge CB3 0HA\hfill\\
$^3$ Present address: Instituto de F\'\i sica, Universidade Federal do Rio Grande do Sul, Porto 
Alegre, BRASIL\\
$^4$ Department of Physics and Astronomy, Johns Hopkins University,
3400 North Charles St, Baltimore MD21218, USA}

\date{Submitted 1997 March 7, Accepted 1997 December 19}
\pagerange{\pageref{firstpage}--\pageref{lastpage}}
\pubyear{1998}

\def\Msun{\hbox{$M_{\odot}$}}
\def\Hunits{\hbox{$\rm km\,s^{-1}\,Mpc^{-1}$}}
\def\phiunits{\hbox{$h^3\,\rm \hbox{Mpc}^{-3}$}}

\def\HST{{\it HST\/}}
\def\etal{{\it et~al.}}
\def\gs{\mathrel{\lower0.6ex\hbox{$\buildrel {\textstyle >}
 \over {\scriptstyle \sim}$}}}
\def\ls{\mathrel{\lower0.6ex\hbox{$\buildrel {\textstyle <}
 \over {\scriptstyle \sim}$}}}
\def\micron{\hbox{$\mu\rm m$}}
\def\sbunits{\hbox{$\hbox{mags}/\hbox{arcsec}^{-2}$}}

\def\PKS{\hbox{$\log P_{KS}$}}
\def\XS{\hbox{$XS$}}

\begin{document}

\label{firstpage}

\maketitle

\begin{abstract}

The excess numbers of blue galaxies at faint magnitudes is a subject of
much controversy. Recent {\it Hubble Space Telescope} results has
revealed a plethora of galaxies with peculiar morphologies tentatively
identified as the evolving population. We report the results of optical
spectroscopy and near-infrared photometry of a sample of faint
\HST\ galaxies from the {\it Medium Deep Survey} to ascertain the
physical properties of the faint morphological populations.  We find
four principal results: Firstly that the population of objects
classified as `peculiar' are intrinsically luminous in the optical
($M_B\sim -19$). Secondly these systems tend to be strong sources of
[OII] line luminosity. Thirdly the optical-infrared colours of the
faint population (a) confirm the presence of a population of {\em
compact\/} blue galaxies and (b) show the stellar populations of
Irregular/Peculiar galaxies encompass a wide range in age. Finally a
surface-brightness comparison with the local galaxy sample
of Frei \etal\ shows that these objects are not of
anomalously low surface brightness, rather we find that all
morphological classes have evolved to a higher surface brightness
at higher-redshifts ($z>0.3$).

\end{abstract}


\section{INTRODUCTION}

The use of the {\it Hubble Space Telescope} (HST) 
has revolutionised the
study of high-redshift galaxy populations. It is well known that counts
of galaxies in blue passbands increasingly exceed no-evolution
predictions at faint magnitudes ($B>20$) (Ellis 1997).  Extensive
redshift surveys have been undertaken of these objects selected in $B$
(Broadhurst \etal\ 1988, Colless \etal\ 1990, Glazebrook \etal\ 1995A,
Ellis \etal\ 1996), $I$ (Lilly 1995) and $K$ bands (Glazebrook
\etal\ 1995B, Cowie \etal\ 1994, 1996) which have been used to construct the
respective luminosity functions as they evolve with redshift. At short
wavelengths, where the evolution implied by the counts is strongest,
there appears to be an increase in the space-density of galaxies at
luminosity $M_B\sim -19$ (we use $H_0=100 \, \rm km\, s^{-1} \,
Mpc^{-1}$),  over $0<z<0.5$ (Ellis \etal\ 1996).  At longer $I$ and $K$
wavelengths, the excess appears to occupy fainter portions of the
luminosity function (Glazebrook \etal\ 1995B, Cowie \etal\ 1994, 1996).

More recently the availability of deep HST data has started to reveal
the morphological character of these faint galaxy populations.
Glazebrook \etal\ (1995C) and Driver \etal (1995) published the first
morphologically split number-magnitude counts to $I=22$ based upon
human classification of faint {\it Medium Deep Survey} data. They
found two principal results: firstly the counts of elliptical and
spiral galaxies matched closely the predictions of a no-evolution model
provided a high local normalisation ($\phi^*=0.03\phiunits$) was used.
Secondly they found the counts of irregular/peculiar galaxies ({\it
i.e.} those lying outside the standard Hubble sequence) rose much
faster than the no-evolution prediction, and it was this steep rise
which appeared to account for the previously known faint blue galaxy
excess.

An obvious problem with this type of analysis is the subjectivity of
human classification and possible systematic effects on the observed
morphology due to cosmological dimming in surface brightness and the
shift of the observed bandpass towards the blue. This was investigated
by Abraham \etal (1996A) who used an objective scheme based upon
central-concentration and asymmetry image parameters to classify
galaxies. To calibrate the systematics they used a sample of CCD images
of nearby galaxies and simulated their appearance to HST at redshifts
$0<z<1$. This confirmed the earlier results of the steep rise in the
number-magnitude counts
of morphologically peculiar system. This trend has now been
shown to extend to $I=25$ (Abraham \etal\ 1996A) using the {\it Hubble
Deep Field\/} data.

It now seems well established that the fraction of peculiar systems is
enhanced at intermediate redshifts; however number-magnitude
counts are rather a crude tool and insensitive to
subtle evolutionary trends in the galaxy  populations.
The next obvious step is to
correlate the morphological properties derived from \HST\ with the
traditional ground-measurable properties such as redshift, luminosity,
line strength and colour to try and establish the physical properties
of this peculiar population and compare it with the more regular
population.  This requires redshifts of the galaxies and to date most
of the large-area \HST\ morphological data has come from the {\it
Medium Deep Survey}, or MDS, (Griffiths \etal\ 1994). Since these are
semi-random parallel fields near other targets of interest the faint
galaxies usually have no prior observations.

To remedy this we have carried out a ground-based observational campaign
of optical spectroscopy and infrared photometry of selected MDS fields
in order to further elucidate the nature of the faint irregular and
normal galaxies. We report here on the results of our observations. The
plan of this paper is as follows: in section 2 we detail our
ground-based observations and the data reduction. Section 3, analyses
the luminosities and line strengths of the faint galaxies, section 4
looks at the optical-infrared colours and section 5 covers the 
surface-brightness distributions.  Finally we summarise our
conclusions in section 6.

\section{GROUND-BASED OBSERVATIONS}

The observations on which this paper is based are divided into two parts:
firstly optical spectroscopy at the 4.2m William Herschel Telescope
(WHT) on La Palma and secondly infrared photometry obtained at the UK
Infrared Telescope (UKIRT) on Mauna Kea, Hawaii.

\subsection{Sample selection}

For our observations we wished to select a flux-limited sample of MDS
galaxies of depth comparable to the deepest existing redshift surveys,
{\it i.e.} covering the redshift range $0<z<1$ (Glazebrook
\etal\ 1995A, Lilly \etal\ 1995). Our objects were drawn from the deep
MDS subset described initially in Glazebrook \etal\ 1995C and published
in full in Abraham \etal (1996A). We excluded the objects classified as
stars on the \HST\ images. The galaxy counts are complete to $I(F814W)
= 23$ and the morphological classification is reliable to  $I(F814W) =
22$. F814W is close to the Cousin's $I$-band and so by selecting all
galaxies with $I(F814W)<22$ we will obtain a sample with a mean
redshift of $\sim 0.5$ (Lilly \etal\ 1995).

\subsection{WHT spectroscopy}

The spectroscopic observations were obtained using the LDSS2 multislit
spectrograph during two runs: 4--6 November, 1994 and 25--27 May,
1995.  A total of 4 clear nights were obtained. A full description of
the LDSS2 spectrograph can be found in Allington-Smith \etal (1994).

An LDSS2 mask was made for each of the MDS fields observed. LDSS2 has a
$10'$ field for multislits, however since the \HST\ MDS fields were
only $\sim 3'$ in size only $\sim$8--13 slits could be put on target
galaxies per mask.  A total of 5 masks were observed containing a total
of 52 galaxies with an accumulated exposure time of 3-4 hours per
mask.  The observational procedure and data reduction were otherwise
identical to that for the LDSS2 $B<24$ redshift survey described in
Glazebrook \etal\ 1995A. ($I<22$ and $B<24$ are broadly equivalent in
terms of typical $S/N$ and  mean redshift.)

The spectral identifications are given in Table~1. The object IDs are
the same as given in Table~1 in the large catalog paper of Abraham
\etal (1996A) which has been cross-referenced for magnitudes, colours,
classifications and other \HST\ image parameters.
We note that several
of these fields were omitted from the original Abraham \etal\ paper due
to the galactic latitude cut to avoid crowded fields. The crowding was
subsequently found not to affect the A/C analysis or the photometry.
These A/C values have been redone and are included in Table~1,
except in two cases where the images were too close to other bright
objects for the A/C analysis. 

The actual spectra are broadly similar to the range of types and $S/N$
shown in Figure~1 of Glazebrook \etal\ and are not reproduced here. The
use of the quality parameter $Q$ for redshift identification confidence
is the same as in Glazebrook \etal. and should have a similar
reliability.

\begin{figure}
\psfig{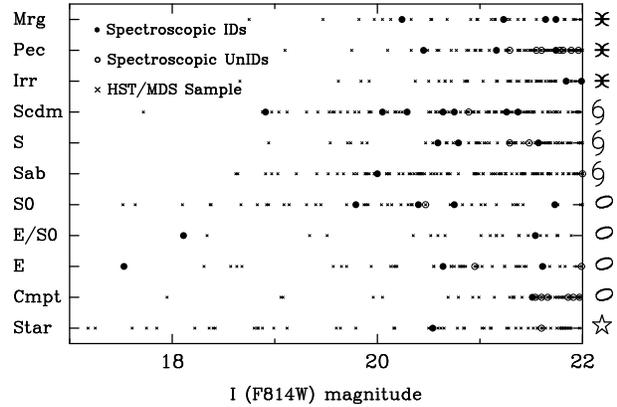}
\caption{
The morphological type vs $I$ magnitude for
the objects in the photometric sample and the spectroscopic sample.
The classifications are those of RSE from Abraham \etal (1996A) and the
key to the right shows the symbols used for these in later figures.
}
\end{figure}

A total of 31 spectra were identified, one of which was a star
misclassified as an S0.  The typical completeness of each mask was
$\sim 50$--70\%, the overall completeness was $60\%$. While this was
somewhat lower than the earlier LDSS2 survey  we believe that was due
to insufficient $S/N$ on the fainter objects as most of the
unidentified objects are concentrated in the range $21.5<I<22$. If
we just consider the brighter $I<21.5$ sample the completeness is 78\% out
out of a total sample of 27 objects. In the following discussion we
identify (by symbol size) the bright sample in the Figures and indicate
the effect of identification completeness on our inferences.
We believe our sample is robust enough for the 
scope of the analyses we will present below
(primarily continuum and line luminosities)
since the number-redshift distribution at $I\ls 22$ is already
well established (Lilly \etal). The most 
important consideration is to ensure the identifications cover the entire
distribution of morphological types revealed by \HST. This distribution
is shown in Figure~1 which demonstrates that the identifications
does indeed covers the range of types. They also cover
the whole of the asymmetry-concentration diagram presented in
Abraham \etal\ (1996A) as an alternative but equivalent measurement of
the morphology. Thus we conclude our spectroscopic sample is 
is a reasonable, albeit small, basis for looking at the physical
parameters as a function of morphology. The number-redshift distribution
is discussed below in Section~3.

\subsection{UKIRT photometry}

The infrared observations for this project were collected on two
observing runs: December 4--6 1994 and May 4--6 1995.  The data was
taken using the infrared camera IRCAM3 which is a $256 \times 256$
indium antimonide (InSb) array. A plate scale of $0.286''$ was used
giving a field of view of $73''$ which is approximately the same as one
\HST\ chip in the group of three in the Wide Field Camera images.

In each field we observed 1--3 \HST\ chips, choosing the chips and the
exact centers in order to maximise the number of $I<22$ galaxy targets.

As the night sky varies considerably on the time-scale of 15--30 minutes
in the $K$-band we made individual exposures of 2 minutes (each consisting
of 12 frames of 10 seconds exposure averaged together), 
making inter-field offsets
and reconstructing the sky flat-field by median filtering in groups of 8.
For the December run we offset the telescope between the \HST\ chips modulo
a $\pm 30$ pixel dither pattern to attempt to get the best possible flatfield.
However this tended to upset the guiding as often the guide star would
be offset past the field dichroic causing a shift of a few arcseconds in
its optical image. Because of this during the May run we simply dithered
at each \HST\ chip position separately. The guiding was greatly improved
and no significant difference was obtained in the quality of the
flatfield.

We had four clear nights in total over the two runs and obtained a
K-band limit (3$\sigma$ per pixel in a $2''$ diameter aperture) 
of $K\simeq 21.5$ in 11 \HST\ fields covering 21 \HST\ chip positions.
with total on-target exposure times ranging from 3000\,s (in good
conditions) to 10\,000\,s (in conditions of moderate extinction). At
this limit we detected virtually all of our $I<22$ targets in this
field. To obtain close to total $K$-magnitudes we used a $6''$ diameter
aperture (matching our corrected \HST\ apertures).  In the sample
172 of the 218 observed galaxies had photometry $\delta m_K <0.2$ mags
(roughly equivalent to $5\sigma$ detections)
and 
202 of the 218 observed galaxies had photometry $\delta m_K <0.5$ mags
(roughly equivalent to $2\sigma$ detections),
the rest
being too faint in $K$. We use the latter as our cutoff for our $I-K$
photometry --- since we doing photometry at sky positions determined from
our $I$-band images object {\em detection} is not an issue and so we
can go deeper into the noise.
These $K$-magnitudes are given in Table~2 along with coordinates and
A/C values from Abraham \etal. Note the A/C values of Abraham \etal\
have only be tabulated for $I<22$, where their analysis stops. Additionally
a few extra objects with $I<22$ were also omitted from the A/C analysis because
the authors quite conservative in rejecting objects too close to the field
edge, too close to other objects or contaminated by weak cosmic ray
events or diffraction spikes.

\section{THE LUMINOSITY OF FAINT GALAXIES}

It has often been argued that the faint blue galaxy
population could be a result of underestimating the
number of local low surface-brightness or low-luminosity
galaxies (e.g. McGaugh \etal\ 1994). These are then
uncovered by the deep imaging surveys which have
a lower surface brightness limit. However since the
excess population has a similar $n(z)$ to the no-evolution
prediction (Glazebrook \etal\ 1995A) it can be inferred 
statistically that this is probably not the case: since
the peak in $n(z)$ occurs for $L^*$ galaxies (where $L^*$
is the characteristic Schechter luminosity in the no-evolution
prediction) then the typical luminosity of the excess population
must be $\sim L^*$. This has been confirmed recently by 
luminosity function analyses of much larger samples (Ellis \etal\
1996) which show an increase in space density over $0<z<0.5$ 
at $M_B \simeq -19$ (for $H_0 = 100\ \Hunits$).

With the sample we present here we are able to look at the
luminosities of individual galaxies directly, in order to 
see how it depends on
morphology.

\subsection{$B$-band Absolute Magnitudes}

\begin{figure}
\psfig{figure=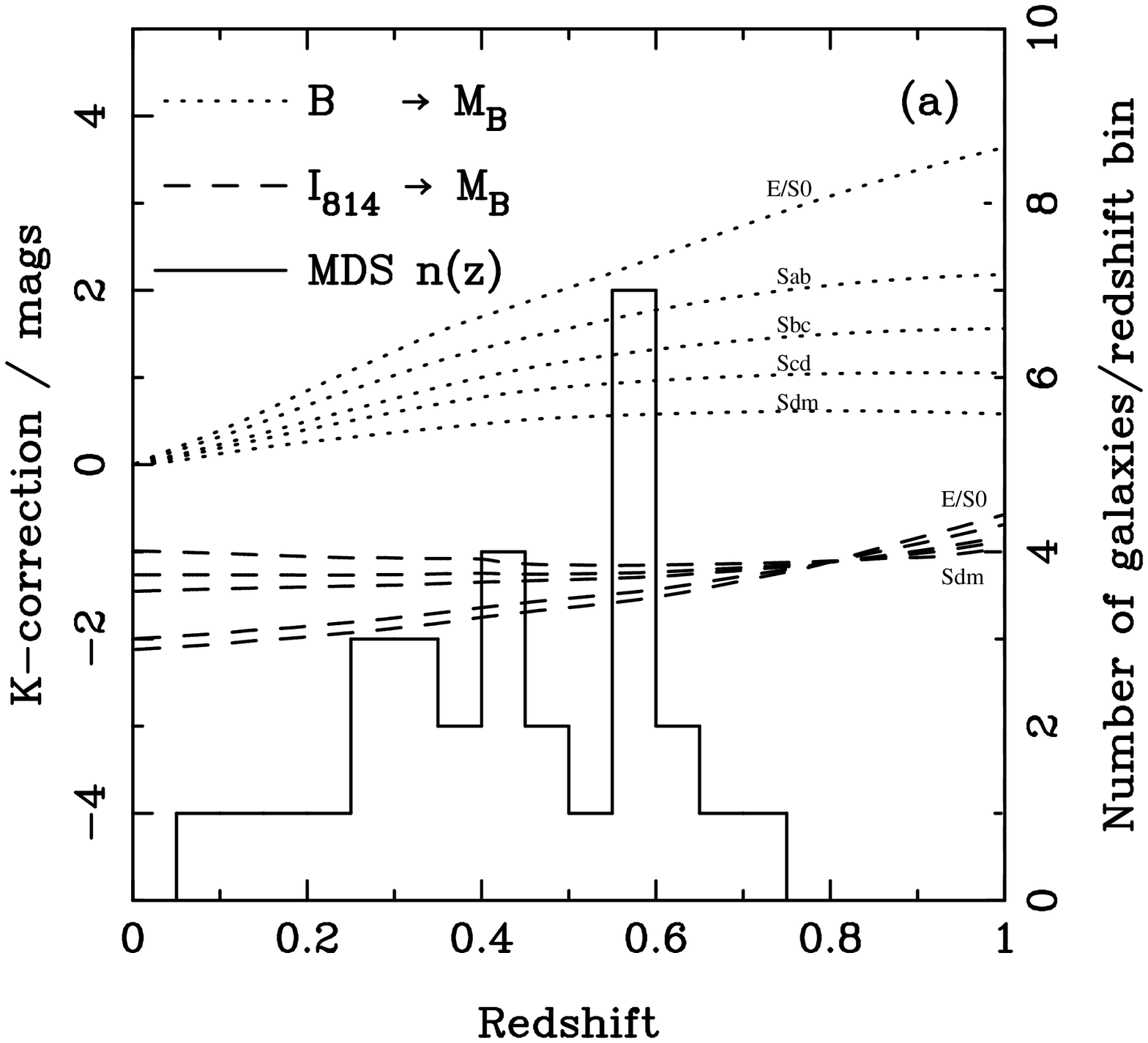,width=8cm,angle=-90}
\medskip
\psfig{figure=mdsphys_fig2b.ps,width=8cm,angle=-90}
\caption{ (a) The redshift distribution $n(z)$ of our data
together with the K-corrections ($I$(observed) to $B$(rest) and
$B$(observed) to $B$(rest)) for our E--Irr SEDs. (b) The
redshift distribution broken down by morphological type.
}
\end{figure}

Since the observed $I$-band at $z\sim 0.8$ is 
close to the restframe $B$-band,
the K-correction is close to constant for all redshifts.
This is shown in Figure~2(a) which shows the $n(z)$
distribution of our sample overlayed with the $I$(observed) to
$B$(rest) and $B$(observed) to $B$(rest) K-corrections using our
spectral energy distribution (SED) templates of local Hubble types
(those of Kennicutt 1992).  The
magnitude-redshift distributions of our morphological types is shown in
Figure~2(b).

It can be seen that at the median redshift of our sample ($z=0.43$)
the range of K-correction over the SEDs is only half that
it is in the $B$ band. For $z<0.8$ the $I$-band corresponds
to rest frame $V$ through $R$ so there is no dependency on the 
uncertain near-UV continua of nearby galaxies and our correction gets
{\em better at the higher redshifts}. Finally as we can assign
physical morphological types from our \HST\ images we can assign an
{\em appropriate} K-correction from the corresponding spectral
type. Even a gross error on the scale of several spectral types (e.g.
elliptical/spiral or spiral/Irr) would correspond to an error in
$M_B$ of only 0.2 mags. 
Thus we believe our K-corrections are much more robust than 
is usually the case and we can expect or $M_B$ values to be
limited by the accuracy of the photometry of the original HST data.

\begin{figure}
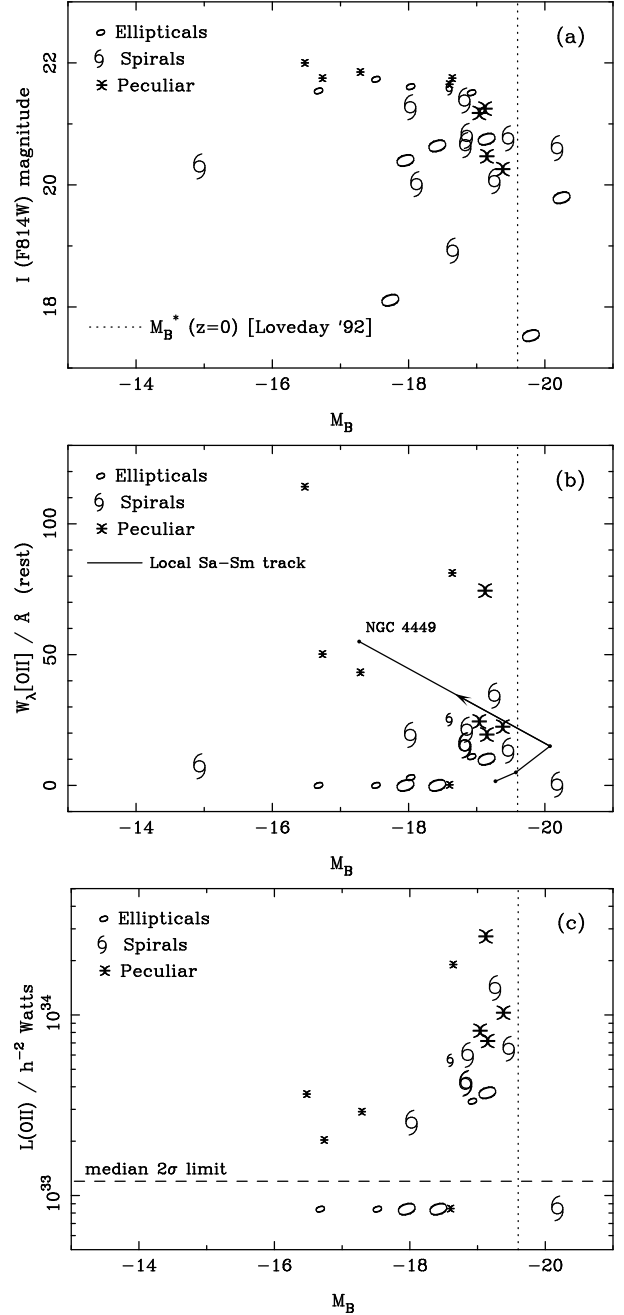

\psfig{figure=mdsphys_fig3a.ps,width=8cm,angle=-90} \smallskip
\psfig{figure=mdsphys_fig3b.ps,width=8cm,angle=-90} \smallskip
\psfig{figure=mdsphys_fig3c.ps,width=8cm,angle=-90} \smallskip
\caption{ $B$-band continuum and [OII] emission line luminosities. The
bright objects ($I<21.5$) are plotted with the large symbols. (a)
$B$ luminosity and $I$ (F814W) 
magnitude. (b) [OII] equivalent widths derived from the
WHT spectra vs $B$ luminosity ($20<I<22$), objects with no measurable
[OII] ($W_\lambda<2$--4\AA) are plotted at $W_\lambda=0$).
(c) Derived [OII] vs $B$ luminosity ($20<I<22$). Objects with no measurable
[OII] are plotted below the dashed line. }
\end{figure}

These $M_B$ values are plotted in Figure~3(a). For a magnitude selected
sample we expect the luminosities to lie near $L^*$ --- this is indeed
the case. {\em It is immediately apparent that the luminosities of the
peculiar systems are very similar to those of the elliptical and spiral
galaxies.} While {\em some} may be dwarfs (all of which are
the fainter $I>21.5$
objects) many have luminosities as bright at
$M_B=-19$. Thus we confirm directly what previously could only be
inferred statistically. This implies that any model which tried to
explain the excess population via local dwarfs {\em must include some
degree of luminosity evolution}.

\subsection{[OII] luminosity}

Since we have spectra we can also look at the line emission of these 
objects. We consider the faint-end slice with $20<I<22$ hereafter
to exclude the brightest, low-redshift objects.
Our spectra were unfluxed but we can directly measure equivalent widths,
this is shown in Figure~3(b). All of the objects had spectral windows
including the [OII]-line location, the strength of this line is an
indicator of star-formation. (Kennicutt 1992).

The previous deep
redshift surveys (Broadhurst \etal\ 1988, Colless \etal\ 1990,
Glazebrook \etal\ 1995A) established that the [OII]
equivalent width distribution showed a high-end tail, not seen
in local samples (Kennicutt 1992). It can be seen from
Figure~3(b) most of the peculiar galaxies have much higher equivalent
widths than the other galaxies, in a region comparable to local
starburst galaxies such as NGC 4449 (Kennicutt 1992). One cautionary
note is that most of these have $I>21.5$ where the incompleteness is
high, obviously our redshift identification would be easier for
strong [OII] emitters. Nevertheless it seems fair to conclude the
high-EW tail (where we expect the completeness to be highest)
is dominated by the morphologically peculiar systems.

Using the equivalent widths and the $M_B$ values we can estimate the
[OII] luminosity which is plotted in Figure~3(c). 
The star-forming $L\sim L^*$ galaxies have $L(OII) \sim 10^{34}
h^{-2}\,\rm W$. [OII] can only be used as a crude estimate of SFRs,
however it is useful to try and estimate this quantity to ascertain the
significance of the star-formation in these system.

If we convert use the conversion value from Kennicutt $10^{34}
h^{-2}\,\rm W$ comes out as 20 $\Msun\rm\,yr^{-1}$ (for $h=0.5$). 
Kennicutt assumes an extinction (1.0 mag at $H\alpha$) typical of local
spirals and solar metallicity. If the metallicity was reduced to
20\% solar the luminosity would be doubled, thus for 
blue and metal-poor systems the corresponding SFR
could be reduced
by a factor of up to 5 --- however the \HST\ data shows some of
them {\em are} morphologically spiral. This is quite a large star-formation
rate which is only found in giant Sc galaxies in the local
Universe --- it is enough to form a
$10^{11}\Msun$ galaxy in 5 Gyr which is $\sim$ the time since $z=0.5$
so it seems clear that we are seeing a major epoch of star formation in
these galaxies unless the bursts are very brief. In the spiral case
at least this seems unlikely as the galaxy would still be visible when 
quiescent and 
nearly all of the spiral and irregular galaxies show this
star-formation.

For the less luminous systems ($M_B>-18$) we see that the star-forming
galaxies have higher equivalent widths. This implies the amount of
star-formation per unit $B$-band light is higher. Finally we note that
though most of the objects classified as `ellipticals' are quiescent
(implying most of the star-formation occurred at $z>1$), many of them
show significant [OII] emission indicating star-formation activity. We
return to this significant point in the next section.

\section{THE COLOURS OF FAINT GALAXIES}

A second method of probing the star-formation in faint galaxies
is via the changes in the integrated colours from their stellar
populations. The most sensitive indices come from the long baseline
provided by optical-infrared colours; the optical light is easily
boosted by a handful of young OB stars radiating in the rest
frame UV--$B$ range while the $\sim 2 \micron$ light comes from
older well-established stellar populations. 

With a sample of 218 galaxies with \HST\ images and $K$-band
magnitudes we have sufficient numbers to construct the colour
distributions broken down according to morphological type 
and compare them with a non-evolving and full spectral-synthesis
predictions.

The colour baseline in our data is provided by the $I-K$ (where $I$ is
F814W from \HST) colours --- their histogram split by \HST\ morphology
is shown in Figure 4. This is, of course the $I-K$ distribution for an
$I<22$ selected sample however it is a straight-forward matter to
calculate the no-evolution prediction for this from a luminosity
function prescription. As said before selecting in the $I$ approximates
a local $B$-selected sample.

Current observational datasets on faint galaxies are quite extensive
--- evolutionary modelers must attempt to fit number-magnitude,
number-redshift, number-colour, colour-colour and colour-magnitude
distributions in bands from $U$ through $K$. Clearly it is too
large a task to reproduce here. Instead we take an existing 
state-of-the-art model as our reference, and see how it does when
the $I-K$ distribution is split up by morphology.

For our reference evolutionary model we use the prescription of
Pozzetti, Bruzual and Zamorani (1996 --- hereafter `PBZ') who 
construct Pure Luminosity
Evolution (PLE) models representing each Hubble type with increasingly
longer star-formation times for later types. Additionally they
introduce a population of `very Blue' galaxies (vB) which are `eternally
young' by which they mean representing the class, at all epochs, by the
SED of a galaxy undergoing a constant star-formation rate (SFR) at age
0.1 Gyr.  This is intended to represent a real galaxy population which
is `cycling', i.e. galaxies bursting with star-formation, fading and
being replaced by others. Such a arbitrary population has also be
introduced by others --- e.g. Gronwall \& Koo (1995), though there
are problems with this approach (see below).

With this extra population and using a high-normalisation
luminosity function (see Glazebrook \etal\ 1995C for a discussion of
normalisation) they match with various degrees of success 
the number-magnitude-colour-redshift distributions mentioned above.
In particular the faint counts and colours ($b_j-r_f$ and $B-K$) seem 
well reproduced to $b_J=24$, though their predicted number-redshift
distributions at $B=24$ see a $z>1$ tail not seen in the data
of Glazebrook \etal\ 1995A. (However there is now evidence from
Cowie \etal 1996 that there may well be such a tail and that
Glazebrook \etal\ incompleteness was biased to $z>1$).

For our modeling we use PBZ's prescription of the luminosity function
which is in turn based upon that of Efstathiou \etal\ (1988).
Following Zucca \etal\ (1994)
PBZ argue that the more recent determination of Loveday \etal\ (1992)
is deficient in faint early-type galaxies. The main change we
make in our modeling is to use the newer Bruzual \&
Charlot (1996) spectral synthesis code (`BC95') which is improved relative to
the Bruzual \& Charlot (1993) code (`BC93') used by PBZ. 
Following PBZ a Gaussian error function with
$\sigma=0.15$ mag is applied in the colour-$z$ plane before deriving
the colour distribution. This is close to the mean $I-K$ error of
our sample ($0.14$ mags). (Even if this was allowed to be bigger 
to allow for the worst case errors (0.2--0.3 mags) the extra smoothing
has no appreciable effect on the model curves below). We have
checked through PBZ's number-colour-magnitude-redshift results
with the 1995 code and our software and find no significant change. When
splitting by morphology we make the following correspondence between
our physically classified types and the spectral types of Table~1 in PBZ:
E/S0 $\Rightarrow$ E/S0, Spiral $\Rightarrow$ Sab, Sbc, Scd and
Irregular/Peculiar $\Rightarrow$ Sdm, vB.

\begin{figure*}
\vspace*{4cm}
\psfig{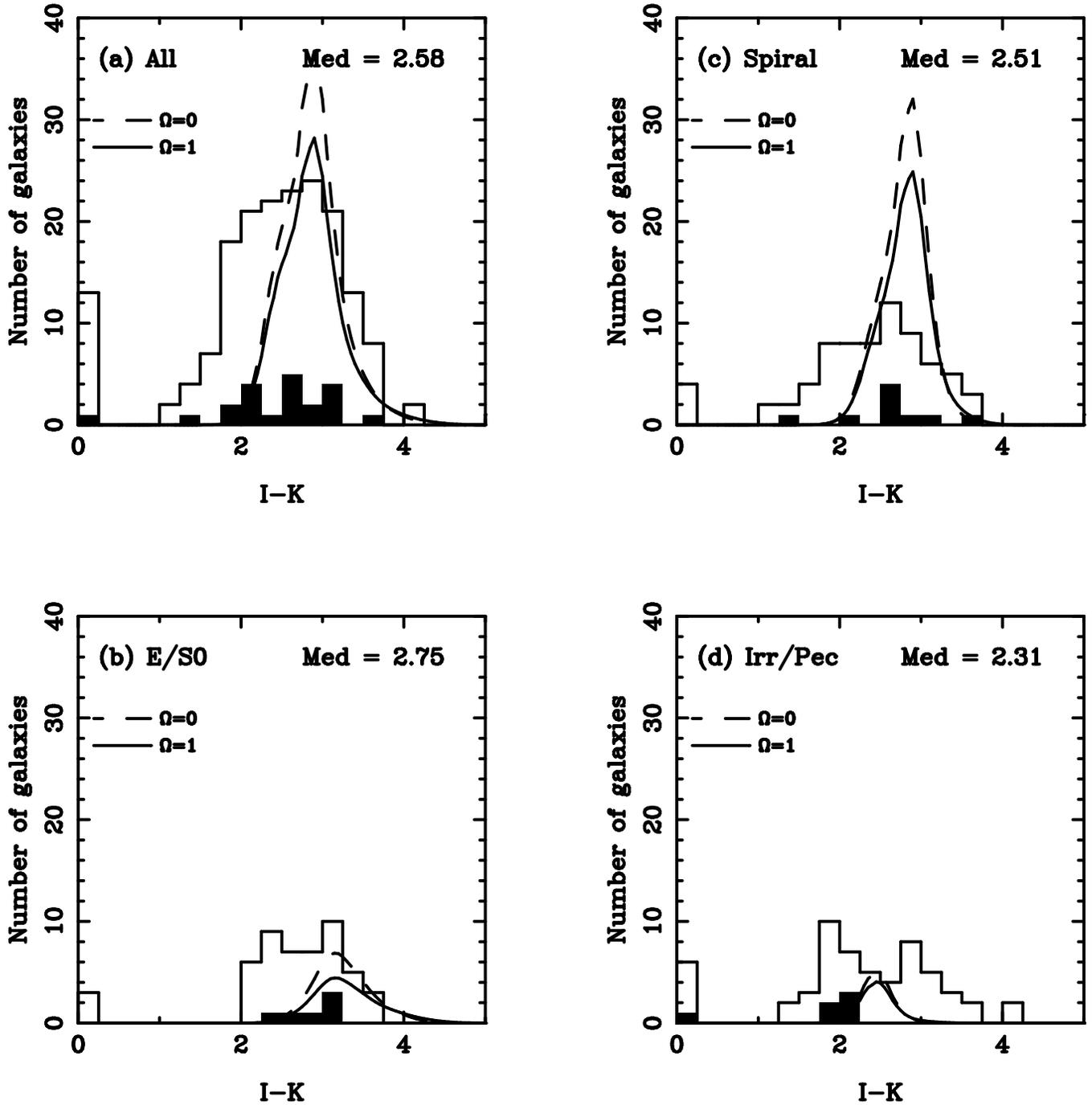} \smallskip
\caption{$I-K$ colour distributions compared with no-evolution
predictions based upon template SEDs (see text for details). 
The open histogram is all the galaxies, the solid histogram is just
those with measured redshifts in our spectroscopic sub-sample.
The median data colour is given in the top
right of each panel. Finally the extreme left-hand bin
in each histogram corresponds
to all galaxies which are too blue to have accurate $K$-band photometry.}
\end{figure*}

Firstly in Figure~4
we show the prediction from no-evolution with
our simple SED templates (note we do not include any `very Blue
population' yet). As a quantitative measure of the significance
of differences between the data and model distribution we use the 
Kolmogorov-Smirnov
test, this is shown in Table~3. When the two differ with more
than 99\% confidence $\PKS < -2$. (Note: to do this we need to
ignore the $K$ non-detections, {\it i.e.} the leftmost bins in
the histogram figures. However these only represent 7\% of objects so
we believe the following conclusions are robust.) Table~3 also give the `excess'
parameter (\XS) --- {\it i.e.} the ratio of the number of galaxies observed
to the number predicted.
Panel (a) of Figure~4 reproduces the known 50--100\% 
(dependent on $\Omega$) excess of faint galaxies at $I=22$, and it
can be seen that most of the excess population is indeed blue.
Breaking down by morphology and inspecting the figures and \PKS\
parameters it is clear that: (a) there is a significant
excess of blue `ellipticals', (b) the spiral distribution has a
bluer median colour and (c) while many of the Irr/Pec galaxies
are indeed very blue many of them have more normal, older
colours indicating that they do not represent a simple `very
blue' population. We have also examined the effect on the
colour distribution
of using the late-age PBZ models as no-evolution SEDs and find
the elliptical predictions become $\sim 0.2$ mag redder
and the spiral$+$Irr prediction becomes $\sim 0.2$ mag {\em bluer}.
This reflects the accuracy with which the PBZ models match the
data at late times. We have also carried out a study of the effects
of metallicity using an early version of the `BC96' code (Bruzual
\& Charlot in preparation). As expected this was small --- varying
the metallicity from solar to 20\% of solar makes the $I-K$ colours
only $0.2$--$0.6$ mag bluer over $0<z<1$.

\begin{figure*}
\vspace*{4cm}
\psfig{figure=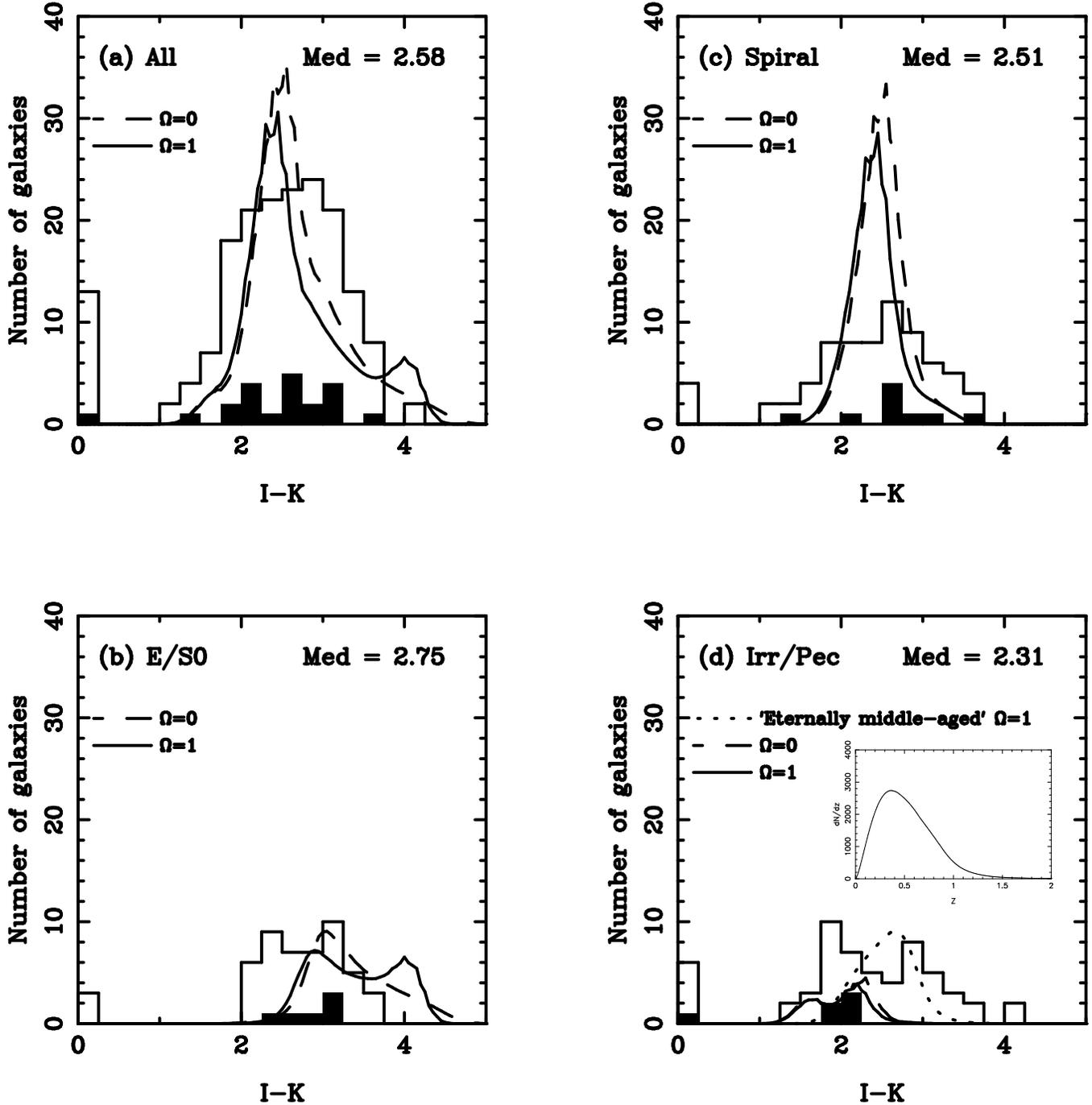,angle=-90} \smallskip
\caption{$I-K$ colour distributions compared with PLE
predictions based upon the PBZ models (see text for details). 
The open histogram is all the galaxies, the solid histogram is just
those with measured redshifts in our spectroscopic sub-sample.
The median data colour is given in the top
right of each panel. In the last panel an additional `eternally 
middle-aged' model is plotted with an arbitrary normalisation.
Finally the extreme left-hand bin in each histogram corresponds
to all galaxies which are too blue to have accurate $K$-band photometry.}
\end{figure*}

To investigate this in more detail we plot in Figure~5 the 
PBZ evolving models (now including the `eternally young' galaxies). 
Several points of
interest are apparent:

\begin{enumerate}

\item 
There is a broad agreement in the colour distribution and
normalisation of {\em all} galaxies.
There is still a $\sim 30$--50\% excess at $I=22$ even
with this model including extra `very Blue' galaxies. This
is also seen in the number-magnitude counts (Figure 3(d)
of PBZ) when plotted on an expanded scale. In our data the significance
of this excess is marginal, especially in light of the uncertainties surrounding
the absolute normalisation of the local luminosity function. However
when we look at the breakdown by morphology a more complex picture
emerges:

\item Even with evolution put in there is still an excess of blue
`ellipticals'. Note the evolution for ellipticals in the
PBZ models is close to `passive evolution' but not quite because
ellipticals are slightly better represented by an exponentially 
decaying SFR with
a short e-folding time rather than a single short burst (though
after 5 Gyr there is little difference). 
Inspection of
the images of the blue galaxies (defining them as $I-K<2.5$) does
indeed show them to be compact objects, occasionally with a very weak
disk. These objects make up 36\% of the objects which are classified as
`elliptical' by the compactness criteria in the HST images and 10\% of
all galaxies --- S0 galaxies should have similar
colours to ellipticals and even contamination by Sa galaxies would
only lead to colours $\sim 0.2$ mag bluer.
We provisionally identify the blue compact galaxies as the same
population as the `Blue Nucleated Galaxies' of Schade \etal\ (1995, 1996A) 
who find a similar proportion (14\%).  All
of the [OII] emitting `ellipticals' in Figure~3 (and one extra with
$I<20$) correspond to galaxies with $I-K<3.2$ and the (strongest
$W_\lambda[\rm OII]=42\AA$) is the bluest ($I=18.1$, $I-K=1.9$).
We hypothesise these do not correspond to local ellipticals since the latter
are reasonably accounted for by the red end of the distribution. Note that
PBZ use a faint end slope for their elliptical luminosity
function of $\alpha=-0.48$, if the slope is flattened to $-1.00$
as used in Glazebrook \etal\ (1995C) the primary effect is 
to increase the normalisation of the model curves by $\simeq 30$\%
and bluen the median colours by 0.2 magnitude (due to
the slightly lower mean redshift). This change does not affect these
arguments.

\item The spiral $I-K$ distribution agrees in mean colour and
normalisation with the model predictions, shifting bluewards by $\sim
0.3$ magnitudes compared to the non-evolving SED prediction. However
this shift is smaller ($0.1$ mag) relative to the non-evolving PBZ
prediction so it is not clear how significant this is.  The blue tail
of the distribution is now better matched, though the range of colours
in the data is still slightly broader than the model prediction. The
discrepancy is quite significant. This implies to us that real
spirals have a more complex distribution of star-formation histories,
or of metallicity or extinction, than in these simple single-history models.

\item The Irregular/Peculiar class are the most interesting in that the
show a surprisingly {\em broad} distribution of $I-K$ colours. In the
models they are only represented by galaxies which are very blue.
The model curves in Figure~5(d) show two blue peaks. The first
peak at $I-K=2.2$ is due
to the model Sdm galaxies and the second one at $I-K=1.6$ is due to the vB
class.  In contrast the data extends out to galaxies with $I-K=4$.
Visual inspection of all Irr/Pec galaxies with $I-K>3$ show that they
do indeed belong in this class.  Examination of their position in the
Asymmetry-Concentration diagram of Abraham \etal\ (1996A) shows the
red Irr/Pec galaxies are not
very different from the blue Irr/Pec galaxies. There is
also no large K-correction effects which may be making the colours
redder --- the inset to Figure~5(d) 
shows the model $n(z)$ which is not too different
from our measured redshifts of these galaxies at this magnitude in
Section~2.  This may not be so convincing since the solid histogram in
Figure~5 shows we only succeeded in getting redshifts for the bluer
galaxies. However the large ground-based Canada-France Redshift Survey
(Lilly \etal\ 1995) was selected to $I<22$ and of $\sim 500$ galaxies the
maximum redshift was only $\sim 1.3$. Sdm galaxies can be as red as
$I-K=3.5$ for $1.3<z<2$ but in this case they would have to be 1--2
magnitudes more luminous. The `eternally young' population is so blue
it has $I-K<2$ even out to $z=4$.  We argue that the most likely
explanation is that a simple representation of this morphologically
peculiar population as a population of `eternally young' (or even Sdm
type) star-forming galaxies is over-simplistic.  The redder galaxies can
be better matched with an older population --- this is demonstrated in
Figure~5 by showing the prediction for an arbitrary `eternally middle-aged'
population of age 5 Gyr post-starburst, which gives a much better
match to the spread of colours. While of course this
is an merely illustrative it is clear that the Irr/Pec population
must be made of galaxies with a spread of age (at least 0--5 Gyr)
unless they were very unusual in metallicity or dust. In the latter
case to match the red end of the $I-K$ histogram an extinction of 
$A_I=1.7$ \,mags is required. However since the sample is selected in
the $I$-band applying this amount of extinction to the galaxy
luminosities reduces the $I<22$ space density by a factor of $\sim 10$.
Thus many more galaxies are required to match the counts.
Of course this could in principle by compensated for by making the
underlying luminosity of the dusty galaxies much greater.
Finally we note that between from 5 to 10 Gyr ($\sim$ $z= 0.5$ to
$z=0$) a pure starburst would fade by $<0.2$ mags in the $UBV$ bands so
{\em should} be seen in the local luminosity function in the absence of
other effects (a point explored in more detail by Bouwens \& Silk, 1996).

\end{enumerate}

\section{THE SURFACE BRIGHTNESS OF FAINT GALAXIES}

\begin{figure}
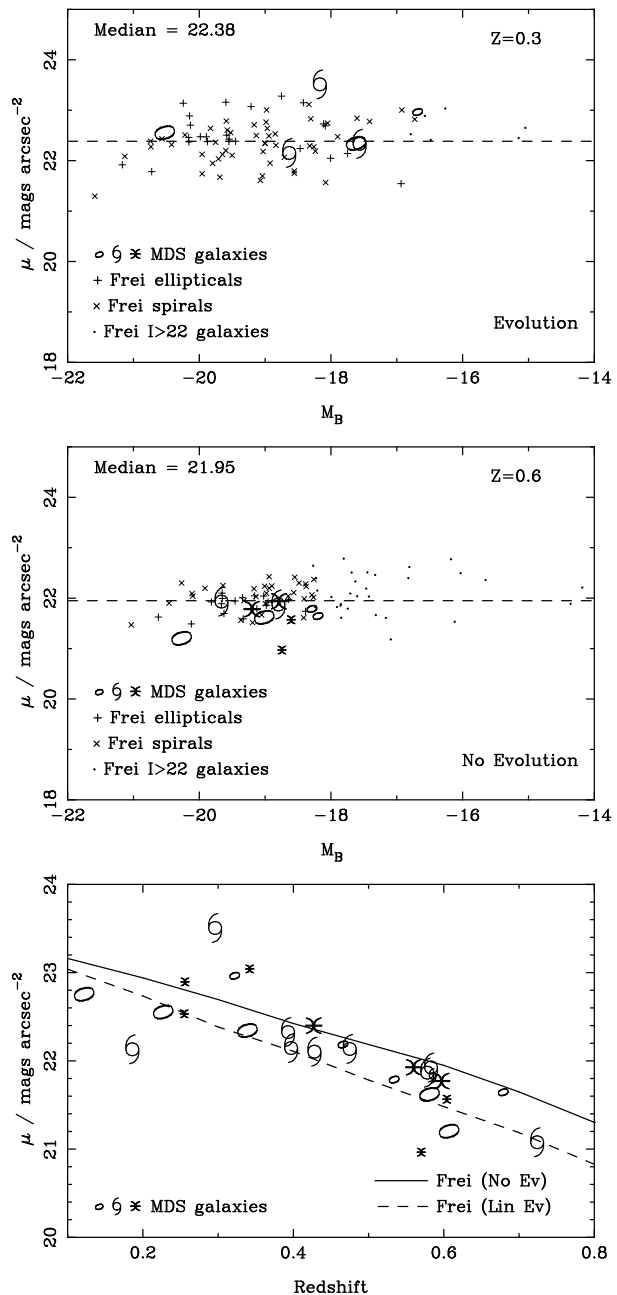

\psfig{figure=mdsphys_fig6a.ps,width=8cm,angle=-90} \smallskip
\psfig{figure=mdsphys_fig6b.ps,width=8cm,angle=-90} \smallskip
\psfig{figure=mdsphys_fig6c.ps,width=8cm,angle=-90} \smallskip
\caption{Differential comparison between the isophotal surface
brightness of MDS galaxies with the artificially redshifted sample
of Frei \etal. Panels (a) and (b) shows examples of the relation
at $z=0.3$ and $z=0.6$  with the MDS galaxies in the
relevant redshift range overlayed (see text)
--- the dashed line shows the median of
the Frei galaxies. Panel (c) shows how the median of the Frei sample
compares with the MDS galaxies versus redshift.}
\end{figure}

A counterpoint to the suggestions that the faint blue galaxy population
may be low-luminosity has been the suggestion that they may constitute a
low surface brightness population.  This is a natural hypothesis
because the deep CCD surveys which uncover the faint blue galaxies also
go to fainter limiting surface brightnesses (McGaugh \etal\ 1994).
However it is difficult to test due to isophotal effects. In a given
survey while surface brightness may be subject to $(1+z)^4$ dimming
with redshift (plus K-corrections) this will also cause the area of a
galaxy above a fixed {\em observed\/} isophote to shrink. Thus only the
inner parts of the galaxy are sampled which can lead to an increase in
{\em isophotal\/} surface brightness.

One approach, as adopted by Schade \etal\ (1995, 1996A, 1996B) is to try and fit
theoretical profile models to the galaxy images --- once the central
surface brightness and scale length is fitted a total magnitude for the
model can be calculated. Of course this requires that the galaxy profile
follows a simple form. 

In our analysis of the MDS data we adopt an orthogonal approach based
on the work of Abraham \etal\ (1996A) --- we compute a simple
parameter (the isophotal surface brightness) and compare against
artificially-redshifted local galaxy templates using the same parameter
to allow for the aforementioned selection effects.  The sample we use
is that  of Frei \etal\ (1996), who chose galaxies which were 
bright, well-resolved and covered a wide range of morphological
classes. The luminosities range from $M_B=-21$ to $M_B=-15$, peaking
around $M^*$, thus approximating quite well the range of luminosities
seen in faint magnitude-selected samples (Brinchmann \etal, 1997).

In the artificial redshifting
procedure  a spectral energy distribution is assigned to each part of
the galaxy, it's image in the F814W filter at the appropriate redshift is
computed and the image is binned up and noise is added so it is
simulated as observed with HST.  Optionally we scale the galaxy flux to
crudely allow for a simple linear brightness evolution ($\Delta M = -2
z$).  Finally we measure the mean surface brightness above a fixed
isophote of $I=24.0 \,\sbunits$ which is approximately twice the
typical noise level in the MDS data. This allows us to make a
differential comparison between the MDS galaxies and the local
galaxies as they would be viewed by HST at the same redshift 
with a minimum of modeling uncertainty.

Figures~6(a) and 6(b) shows examples of this at $z=0.3$ and $z=0.6$
(with and without evolution). We plot the $I<24.0\,\sbunits$ isophotal surface
brightness versus absolute magnitude, for Frei and MDS galaxies (the MDS
galaxies plotted are those within $\pm 0.1$ in redshift). 
Empirically we find that at any particular redshift over $0<z<1$ the Frei
relation is well defined by a constant surface brightness plus scatter
over the sample's luminosity range --- this was
expected because the local Frei sample is simply selected from the NGC
catalog and will follow a Freeman (1970) type law. Thus we can represent the
Frei value by a median value for any particular redshift and
evolutionary scenario. Note in the calculation of the median we exclude
Frei galaxies which fall beyond the MDS $I=22$ limit. Thus the calibration
sample exhibits the same bias to more luminous galaxies at higher redshifts
as the MDS. In practice however we found this exclusion
makes no difference to the final result as the $I>22$ galaxies have similar
surface brightnesses (as can be seen in Figures 6(a) and (b)). 

This leads us to
Figure~6(c) which plots the median of the Frei sample against redshift
(both with and without evolution) and compares with the MDS galaxies.
It can be seen that the MDS galaxies are consistently brighter than
their local counterparts for $z>0.3$. Comparing with the arbitrarily evolved local
counterparts (whose evolution amounts to 2 magnitudes at $z=1$) we estimate
that typically the amount of evolution in the MDS galaxies is about 
half this --- i.e. about 1 magnitude by $z=1$. This can also be seen
directly in the $z=0.6$ redshift slice shown in Figure~6(b). This 
conclusion still holds when considering the $I<21.5$ high-completeness
sub-sample and we conclude we are seeing a genuine evolutionary effect
{\em providing the Frei sample is representative of local galaxies}. 

The amount of this brightening is the same as found by Schade \etal\
from their fitting method, like Schade \etal\ we also find the brightening
appears to apply to objects of all morphologies.

\section{CONCLUSIONS}

We conclude:

\begin{enumerate}

\item The faint-blue galaxy excess to $I=22$ is not due to nearby
under-luminous galaxies being revealed by faint surveys, rather
that many of the objects are observed to be close to $L*$. This is 
true for all morphological classes. Thus, for example, we can
not simply explain the excess of peculiar systems by a uniform
population of low-luminosity dwarf galaxies being revealed by deep surveys.

\item A significant component of the blue excess ($\sim 10\%$)
is composed of compact blue objects originally classified as
`ellipticals'. These are provisionally identified with the
`Blue Nucleated Galaxies' of Schade \etal\ (1995, 1996).

\item The red envelope of the population of compact objects at $z\sim
0.5$ accounts for the number of elliptical galaxies we see today.

\item We see tentative evidence for some mild colour evolution
in the population of spiral galaxies though it is not clear
how significant this is given the uncertainty in the models, 
and overall broader range of colours than exhibited by the models.

\item The galaxies in the Irr/Peculiar morphological class can not simply be
represented by a simple population of very young blue galaxies.  Rather
the broad distribution of blue and red colours indicate a range of
ages (0--5 Gyr), if interpreted as stellar populations, or luminous
dusty galaxies with extinctions of up to 3 mags in the $I$-band. We
conclude that models such as those of Pozzetti \etal\ or Gronwall \&
Koo (1995) are too simplistic.  This adds to the other known problem
with these types of model --- the predicted overabundance of
low-luminosity galaxies at $z=0$ (Bouwens \& Silk, 1996). The properties
and evolution of the very late-type galaxies is clearly not
well understood yet.

\item The line-luminosities in [OII] indicate significant amounts of
star-formation is occurring at $z=0.5$, primarily in the late-type
Irr/Pec population but also in the spirals and blue `ellipticals'.

\item The surface brightness relation shows no evidence that any of the
faint morphological populations are of anomalously low surface
brightness.  Rather we confirm the result of Schade \etal\ (1995,
1996A, 1996B), from a completely different non-parametric method
(comparison with the artificially-redshifted Frei \etal\ local sample), of
evidence for about 1~magnitude evolution towards a higher surface
brightness, in all morphological classes, for $z>0.3$. 

\item There is a lack of success in measuring the redshifts for the
redder peculiar systems, so it is difficult to constrain them in any
way. It is entirely possible they may be anomalous in luminosity,
redshift or surface brightness. However it is clear that the peculiar
population is not homogeneous and this must be accounted for in any
realistic evolutionary model. A spectroscopic campaign targeted at 
these red objects would of immense value in understanding the 
population of peculiar objects revealed in faint HST images.

\end{enumerate}

It is obvious that larger samples with spectroscopy and HST imaging are desirable
to further this work; with the advent of large HST imaging programmes in
Cycles 6 and 7 samples of several hundred objects are becoming
available (e.g. Brinchmann \etal. 1997). Also the spectral synthesis
models are being improved (e.g. the `BC96' Bruzual and Charlot code,
in preparation) and it will be possible to include in the models
effects such as varying metallicity. With very deep multi-colour
HST images there
is also the possibility of looking at the star-formation history
of different portions of an individual galaxy (Abraham \etal. 1997).
It is clear that in the next few years a much more 
detailed understanding of
the evolution and properties of the faint galaxy populations will be
achieved.

\section{ACKNOWLEDGEMENTS}

The authors thank Stefan Charlot for his generosity in making his 1995
spectral synthesis code available to us.
This paper is based on observations with the NASA/ESA {\it Hubble Space
Telescope}, obtained at the Space Telescope Science Institute, which is
operated by the Association of Universities for Research in Astronomy
Inc., under NASA contract NAS5-26555. Coordination and analysis of data
from the Medium Deep Survey is funded by STScI grants GO2684.0X.87A and
GO3917.OX.91A.  We also gratefully acknowledge the generous allocations
of telescope time on the UK Infrared Telescope, operated by the Royal
Observatory Edinburgh and the William Herschel Telescope, operated by
the Royal Greenwich Observatory in the Spanish Observatorio del Roque
de Los Muchachos of the Instituto de Astrof\'\i sica de Canarias.  We
also thank the staff and telescope operators of these telescopes for
their enthusiasm and competent support. The data reduction and analysis
was performed primarily  with computer hardware supplied by STARLINK.
All the authors acknowledge funding for this research from the PPARC.

\vfill\eject

\onecolumn

\section{TABLES}

\bigskip\bigskip
{\bf TABLE 1.}\quad The WHT/LDSS2 Spectroscopic Sample.
\bigskip

{\scriptsize

%\documentstyle[]{article} 
%\oddsidemargin=-0.8in\evensidemargin=-0.8in\begin{document}  
%\noindent{\bf Table 1.} The WHT/LDSS Spectroscopic Sample 

\medskip\let\h=\hfil 
\halign{\tabskip=1em 
 #\h   &  \h#\h &   \h#\h &   \h#\h &   \h#\h &   \h#\h &   \h#\h &  # &   #\h &   # &   #\h     \cr 
 MDS ID       &  
$z$ &  
RA    &  
Dec   &  
$I$  &  
$C$  &  
$A$  &  
Q$^1$  &  
Type$^2$ &  
[OII] EW & 
Comment$^3$ \cr 
 \noalign{\vglue 0.1truecm} 
 ubi1-10 & 0.436 & 01 10 00.98 & -02 28 44.9 & 20.64 & 0.50 & 0.03 & 1 & A &  $ 0 \pm 5 $  & H,K,4000\AA\ break,G\cr
ubi1-8 & \hbox{No Id} & 01 10 00.57 & -02 28 28.4 & 21.66 & 0.42 & 0.00 & 4 & --- &  $ N/A $  & Missing?\cr
ubi1-18 & \hbox{No Id} & 01 09 59.94 & -02 28 15.0 & 21.60 & --- & --- & 4 & --- &  $ N/A $  & Missing?\cr
ubi1-24 & 0.065 & 01 09 58.24 & -02 28 05.0 & 20.29 & 0.32 & 0.06 & 1 & E &  $ 7 \pm 10 $  & [OII]?,[OIII],$H\alpha$$+$,[NII],[SII]\cr
ubi1-2 & 0.428 & 01 10 00.56 & -02 27 46.6 & 20.64 & 0.33 & 0.00 & 1 & E &  $ 15 \pm 2 $  & [OII],H,K,Mgb\cr
ubi1-43 & \hbox{No Id} & 01 09 56.88 & -02 27 32.0 & 21.97 & 0.36 & 0.01 & 4 & --- &  $ N/A $  & Missing?\cr
ubi1-31 & 0.432 & 01 09 58.20 & -02 27 18.7 & 20.24  & 0.33 &  0.19 & 1 & EAB &  $ 22 \pm 1 $  & [OII],$H\beta$$+$,[OII],$H\gamma$$+$,H,K,Balmer\cr
ubi1-48 & \hbox{No Id} & 01 09 59.95 & -02 27 04.7 & 20.47 & 0.46 & 0.09 & 3 & --- &  $ N/A $  & Weird!!! BAL AGN???\cr
ubi1-61 & \hbox{No Id} & 01 10 00.50 & -02 26 49.7 & 21.89 & 0.16 & 0.00 & 3 & --- &  $ N/A $  & Weak\cr
ubi1-70 & \hbox{No Id} & 01 10 01.23 & -02 26 30.5  & 21.78 & 0.57 & 0.13 & 4 & --- &  $ N/A $  & Missing?\cr
ubi1-51 & 0.560 & 01 09 58.50 & -02 26 18.7 & 21.16 & 0.27 & 0.14 & 2 & EAB &  $ 24 \pm 2 $  & [OII],HK?,Balmer?,4000\AA\ break\cr
ubi1-68 & \hbox{No Id} & 01 10 01.16 & -02 26 05.6 & 21.60 & 0.26 & 0.08 & 4 & --- &  $ N/A $  & Missing?\cr
ubi1-55 & 0.427 & 01 09 58.87 & -02 25 56.7 & 20.45 & 0.23 & 0.05 & 1 & EA &  $ 19 \pm 2 $  & [OII],$H\beta$$+$?,[OII]?,H,K,G\cr
uim0-11 & 0.597 & 03 55 31.50 & 09 42 15.1 & 21.23 & 0.23 & 0.32 & 1 & E &  $ 74 \pm 2 $  & [OII],[OIII],$H\beta$$+$?,HK?\cr
uim0-10 & \hbox{No Id} & 03 55 31.34 & 09 42 29.0 & 21.55 & 0.19 & 0.06 & 3 & --- &  $ N/A $  & Weak\cr
uim0-9 & 0.679 & 03 55 31.13 & 09 42 41.1 & 21.51 & 0.46 & 0.00 & 2 & EA &  $ 11 \pm 5 $  & [OII],H,K\cr
uim0-1 & 0.339 & 03 55 33.45 & 09 43 01.8 & 20.40 & 0.50 & 0.04 & 1 & A &  $ 0 \pm 4 $  & K,H,4000\AA\ break,G,$H\beta$$-$,Mgb,5268\cr
uim0-18 & \hbox{No Id} & 03 55 29.22 & 09 43 27.9 & 20.89 & 0.20 & 0.17 & 3 & --- &  $ N/A $  & Em line??\cr
uim0-30 & \hbox{No Id} & 03 55 28.49 & 09 43 40.6 & 21.29 & 0.36 & 0.06 & 3 & --- &  $ N/A $  & Weak\cr
uim0-28 & 0.466 & 03 55 28.54 & 09 43 53.8 & 21.73 & 0.49 & 0.02 & 2 & A &  $ 0 \pm 4 $  & H,K,4000\AA\ break,[OII]?\cr
uim0-37 & \hbox{No Id} & 03 55 31.53 & 09 44 12.9 & 21.48 & 0.43 & 0.04 & 3 & --- &  $ N/A $  & Weak\cr
uim0-38 & 0.475 & 03 55 31.52 & 09 44 23.5 & 20.79 & 0.31 & 0.17 & 1 & EAB &  $ 21 \pm 2 $  & [OII],H,K,Balmer,G,$H\beta$$+$?,[OIII]?\cr
uim0-43 & 0.724 & 03 55 34.03 & 09 44 41.4 & 20.59 & 0.53 & 0.05 & 1 & EAB &  $ 0 \pm 2 $  & H,K,G,Balmer,$+$unk 3594$-$\cr
uim0-42 & 0.256 & 03 55 32.00 & 09 44 59.4 & 21.74 & 0.30 & 0.03 & 1 & E &  $ 50 \pm 4 $  & [OII],$H\beta$$+$,[OIII],$H\alpha$$+$,HK?\cr
ueh0-33 & 0.393 & 00 53 20.73 & 12 33 07.0 & 21.26 & 0.33 & 0.06 & 2 & EAB &  $ 19 \pm 3 $  & [OII],HK?,Balmer?\cr
ueh0-35 & 0.578 & 00 53 20.45 & 12 33 20.2 & 21.37 & 0.25 & 0.14 & 1 & EAB &  $ 15 \pm 4 $  & [OII],strong Balmer,4000\AA\ break,[OII]\cr
ueh0-27 & 0.581 & 00 53 23.03 & 12 33 32.1 & 20.75 & 0.54 & 0.02 & 1 & EAB &  $ 10 \pm 2 $  & [OII],HK,4000\AA\ break,Balmer\cr
ueh0-34 & 0.534 & 00 53 21.17 & 12 33 50.1 & 21.61 & 0.42 & 0.00 & 2 & EAB &  $ 3 \pm 2 $  & [OII]?,HK,4000\AA\ break,G,Balmer\cr
ueh0-43 & 0.585 & 00 53 23.68 & 12 34 17.0 & 21.57 & 0.30 & 0.05 & 1 & EAB &  $ 25 \pm 2 $  & [OII],$H\beta$$+$,HK,Balmer\cr
ueh0-54 & \hbox{No Id} & 00 53 21.54 & 12 34 29.8 & 21.96 & --- & --- & 3 & --- &  $ N/A $  & Weak\cr
ueh0-49 & 0.583 & 00 53 23.57 & 12 34 45.3 & 20.75 & 0.32 & 0.05 & 1 & EAB &  $ 13 \pm 2 $  & [OII],HK,G,Balmer\cr
ueh0-51 & \hbox{No Id} & 00 53 22.42 & 12 35 05.3 & 21.91 & 0.36 & 0.00 & 3 & --- &  $ N/A $  & Weak\cr
ueh0-55 & \hbox{No Id} & 00 53 21.18 & 12 35 17.9 & 21.81 & 0.31 & 0.17 & 3 & --- &  $ N/A $  & Weak $-$ maybe z=1.383 MgII,MgI?\cr
usa0-15 & 0.604 & 17 12 19.41 & 33 35 18.4 & 21.74 & 0.42 & 0.20 & 1 & E &  $ 81 \pm 4 $  & [OII],$H\beta$$+$,[OIII]\cr
usa0-93 & \hbox{No Id} & 17 12 27.48 & 33 35 30.1 & 20.95 & 0.65 & 0.11 & 3 & --- &  $ N/A $  & Too faint\cr
usa0-90 & \hbox{No Id} & 17 12 27.76   &  33 35 42.1 & 21.60 & 0.43 & 0.42 & 3 & --- &  $ N/A $  & Bright, featureless\cr
usa0-91 & \hbox{No Id} & 17 12 29.68 & 33 36 19.1 & 22.00  &  0.34 & 0.06 & 3 & --- &  $ N/A $  & Too faint\cr
usa0-69 & 0.296 & 17 12 24.83 & 33 37 03.8 & 20.00 & 0.20 & 0.90 & 1 & EAB &  $ 36 \pm 4 $  & [OII],$H\beta$$+$,[OIII],$H\alpha$$+$,HK,Balmer$-$\cr
usa0-66 & 0.342 & 17 12 22.11 & 33 37 17.7 & 21.84 & 0.24 & 0.08 & 1 & EAB &  $ 43 \pm 18 $  & [OII],$H\beta$$+$,[OIII],HK,G,Balmer$-$\cr
usa0-57 & 0.255 & 17 12 25.88 & 33 36 36.1 & 21.99 & 0.46 & 0.06 & 1 & E &  $ 114 \pm 16 $  & [OII],$H\beta$$+$,[OIII],$H\alpha$$+$,$H\delta$$+$?\cr
ux40-108 & 0.227 & 15 19 37.96 & 23 50 51.7 & 17.53 & 0.64 & 0.04 & 1 & A &  $ 0 \pm 2 $  & H,K,G,Mgb,NaD\cr
ux40-102 & 0.322 & 15 19 39.15 & 23 51 06.5 & 21.54 & 0.46 & 0.01 & 1 & EAB &  $ 0 \pm 23 $  & [OII]?,$H\beta$$+$,[OIII],$H\alpha$$+$,HK,Balmer$-$\cr
ux40-116 & 0.000 & 15 19 40.49  & 23 51 26.89  & 20.54 & 0.82 & 0.01 & 1 & S &  $ N/A $  & M star\cr
ux40-114 & \hbox{No Id} & 15 19 40.49 & 23 51 26.9 & 21.86 & 0.72 & 0.01 & 3 & --- &  $ N/A $  & Featureless\cr
ux40-51 & \hbox{No Id} & 15 19 42.45 & 23 51 57.0 & 21.54 & 0.46 & 0.06 & 3 & --- &  $ N/A $  & Possible MgII,MgI at z=1.202?\cr
ux40-83 & 0.570 & 15 19 43.30 &  23 52 13.2  & 21.64 & 0.75 & 0.04 & 2 & EAB &  $ 0 \pm 1 $  & [OIII]?,HK?,Balmer$-$?\cr
ux40-3 & 0.607 & 15 19 40.62 & 23 52 21.6 & 19.79 & 0.39 & 0.17 & 2 & EA &  $ 0 \pm 1 $  & [OII]?,$H\beta$?,HK?,Mgb?\cr
ux40-19 & \hbox{No Id} & 15 19 41.92 & 23 52 33.3 & 21.99 &  0.35 & 0.04 & 3 & --- &  $ N/A $  & Too faint\cr
ux40-27 & 0.397 & 15 19 42.94 & 23 52 46.2 & 20.05 & 0.28 & 0.17 & 1 & EAB &  $ 34 \pm 2 $  & [OII],$H\beta$$+$,[OIII],HK,Balmer$-$\cr
ux40-7 & 0.186 & 15 19 38.86 & 23 53 02.1 & 18.91 & 0.43 & 0.16 & 1 & EAB &  $ 15 \pm 2 $  & \vtop{\hbox{[OII],$H\beta$$+$,[OIII],$H\alpha$$+$,[NII],[SII],HK,}\hbox{Balmer$-$,G,Mgb}}\cr
ux40-26 & \hbox{No Id} & 15 19 41.29 & 23 53 13.0 & 21.29 & 0.19 & 0.02 & 3 & --- &  $ N/A $  & Too faint\cr
ux40-28 & 0.122 & 15 19 40.85 & 23 53 33.8 & 18.11 & 0.60 & 0.12 & 1 & EAB &  $ 42 \pm 4 $  & \vtop{\hbox{[OII],$H\beta$$+$,[OIII],$H\alpha$$+$,[NII],[SII],HK,}\hbox{Balmer$-$,Mgb,OI(6300),7267$+$}}\cr
} 
%\end{document} 

  \bigskip

  \begin{enumerate}

  \item[1.] Redshift Quality values: $1=$High confidence, $2=$normal confidence,
  $3=$No identification, $4=$No signal in spectrum.

  \item[2.] Type symbols: E$=$Emission lines present, A$=$Absorption lines present, B$=$Balmer
  series present (A star signature).

  \item[3.] In comments: $+=$ emission feature, $-=$ absorption feature.

  \end{enumerate}

}

\vfill\eject
\twocolumn
\bigskip\bigskip
{\bf TABLE 2.}\quad $K$-band UKIRT Photometry.
\bigskip

{\scriptsize

%\documentstyle[]{article} 
%\begin{document}  
%\noindent{\bf Table 2.} The K-band Photometric Sample 

\medskip\let\h=\hfil 
\halign{\tabskip=1em 
 #\h   &  \h#\h &   \h#\h &   \h#\h &   \h#\h &   \h#\h &   \h#\h \cr 
 MDS ID       &  
RA    &  
Dec   &  
$I$  &  
$K$  &  
$C$  &  
$A$  \cr  
 \noalign{\vglue 0.1truecm} 
 u3p0-2 & 07 00 00.68 & 14 09 28.7 & 21.97 & \hbox{No signal} & 0.38 & 0.01\cr
u3p0-3 & 07 00 00.31 & 14 09 30.4 & 22.38 & $ 19.41 \pm 0.21 $ & --- & ---\cr
u3p0-4 & 06 59 58.64 & 14 09 42.2 & 24.90 & \hbox{No signal} & --- & ---\cr
u3p0-5 & 06 59 57.94 & 14 09 27.6 & 21.42 & $ 17.72 \pm 0.05 $ & 0.56 & 0.00\cr
u3p0-7 & 06 59 58.83 & 14 09 35.5 & 99.99 & \hbox{No signal} & --- & ---\cr
u3p0-8 & 06 59 58.20 & 14 09 26.5 & 99.99 & $ 19.90 \pm 0.45 $ & --- & ---\cr
u3p0-9 & 06 59 59.59 & 14 09 09.4 & 21.65 & $ 17.47 \pm 0.04 $ & 0.29 & 0.95\cr
u3p0-10 & 06 59 58.22 & 14 09 23.6 & 23.47 & $ 19.80 \pm 0.33 $ & --- & ---\cr
u3p0-13 & 06 59 59.52 & 14 09 33.0 & 25.35 & \hbox{No signal} & --- & ---\cr
u3p0-14 & 06 59 57.30 & 14 09 55.8 & 24.89 & $ 22.78 \pm 4.25 $ & --- & ---\cr
u3p0-15 & 06 59 58.23 & 14 09 20.1 & 22.38 & $ 19.43 \pm 0.23 $ & --- & ---\cr
u3p0-16 & 06 59 58.70 & 14 09 07.5 & 23.40 & $ 20.40 \pm 0.57 $ & --- & ---\cr
u3p0-17 & 06 59 56.94 & 14 10 22.5 & 21.43 & $ 18.09 \pm 0.06 $ & 0.23 & 0.13\cr
u3p0-18 & 06 59 55.01 & 14 10 07.5 & 20.76 & $ 17.60 \pm 0.04 $ & 0.48 & 0.04\cr
u3p0-19 & 06 59 55.47 & 14 10 19.4 & 19.53 & $ 16.54 \pm 0.01 $ & --- & ---\cr
u3p0-20 & 06 59 55.94 & 14 10 26.6 & 23.51 & $ 22.03 \pm 2.21 $ & --- & ---\cr
u3p0-21 & 06 59 56.19 & 14 10 33.1 & 21.55 & $ 18.67 \pm 0.10 $ & 0.40 & 0.06\cr
u3p0-22 & 06 59 54.83 & 14 10 16.8 & 24.21 & $ 20.36 \pm 0.51 $ & --- & ---\cr
u3p0-23 & 06 59 54.34 & 14 10 16.9 & 22.25 & $ 19.36 \pm 0.20 $ & --- & ---\cr
u3p0-24 & 06 59 54.13 & 14 10 24.4 & 24.23 & $ 20.56 \pm 0.62 $ & --- & ---\cr
u3p0-25 & 06 59 53.94 & 14 10 21.8 & 23.60 & $ 19.99 \pm 0.40 $ & --- & ---\cr
u3p0-26 & 06 59 55.30 & 14 10 51.2 & 20.89 & $ 19.15 \pm 0.15 $ & 0.31 & 0.02\cr
u3p0-27 & 06 59 55.59 & 14 10 56.8 & 22.26 & $ 20.18 \pm 0.43 $ & --- & ---\cr
u3p0-28 & 06 59 53.96 & 14 10 43.9 & 24.22 & $ 20.93 \pm 0.92 $ & --- & ---\cr
u3p0-30 & 06 59 55.63 & 14 09 52.2 & 23.80 & $ 19.64 \pm 0.33 $ & --- & ---\cr
u3p0-31 & 06 59 57.77 & 14 10 33.8 & 21.76 & $ 18.34 \pm 0.08 $ & 0.23 & 0.02\cr
u3p0-32 & 06 59 56.99 & 14 10 26.0 & 22.54 & $ 19.29 \pm 0.19 $ & --- & ---\cr
u3p0-33 & 06 59 55.28 & 14 09 59.1 & 22.24 & $ 18.92 \pm 0.17 $ & --- & ---\cr
u3p0-35 & 06 59 54.22 & 14 10 02.9 & 24.05 & $ 20.36 \pm 0.57 $ & --- & ---\cr
u3p0-36 & 06 59 54.92 & 14 10 43.0 & 28.20 & $ 20.70 \pm 0.61 $ & --- & ---\cr
u3p0-37 & 06 59 54.00 & 14 10 32.8 & 25.28 & \hbox{No signal} & --- & ---\cr
u3p0-38 & 06 59 54.09 & 14 10 37.9 & 24.82 & $ 19.54 \pm 0.24 $ & --- & ---\cr
u3p0-39 & 06 59 54.22 & 14 10 49.2 & 23.59 & $ 19.25 \pm 0.17 $ & --- & ---\cr
u3p0-40 & 06 59 56.20 & 14 10 12.5 & 23.49 & \hbox{No signal} & --- & ---\cr
u3p0-42 & 06 59 53.89 & 14 10 54.9 & 22.63 & \hbox{No signal} & --- & ---\cr
u3p0-46 & 06 59 58.46 & 14 11 08.3 & 20.84 & $ 18.07 \pm 0.13 $ & --- & ---\cr
u3p0-52 & 06 59 57.13 & 14 11 00.2 & 21.90 & $ 17.87 \pm 0.05 $ & --- & ---\cr
u3p0-53 & 06 59 57.92 & 14 10 49.4 & 21.99 & $ 17.93 \pm 0.06 $ & --- & ---\cr
u3p0-54 & 06 59 57.14 & 14 11 04.7 & 99.99 & \hbox{No signal} & --- & ---\cr
u3p0-55 & 06 59 58.69 & 14 10 48.3 & 23.95 & $ 18.61 \pm 0.15 $ & --- & ---\cr
ubi1-1 & 01 10 02.73 & -02 27 52.0 & 20.10 & $ 17.06 \pm 0.03 $ & 0.34 & 0.38\cr
ubi1-2 & 01 10 00.56 & -02 27 46.6 & 20.64 & $ 17.94 \pm 0.07 $ & 0.33 & 0.00\cr
ubi1-3 & 01 10 03.37 & -02 28 25.4 & 20.81 & $ 17.98 \pm 0.06 $ & 0.18 & 0.16\cr
ubi1-4 & 01 10 00.38 & -02 28 05.2 & 22.05 & $ 19.30 \pm 0.21 $ & --- & ---\cr
ubi1-5 & 01 10 00.48 & -02 28 11.2 & 21.94 & \hbox{No signal} & 0.28 & 0.06\cr
ubi1-6 & 01 10 00.03 & -02 28 19.5 & 19.31 & $ 16.90 \pm 0.03 $ & 0.15 & 0.26\cr
ubi1-7 & 01 10 01.75 & -02 28 36.0 & 20.97 & $ 19.02 \pm 0.15 $ & 0.28 & 0.08\cr
ubi1-8 & 01 10 00.57 & -02 28 28.4 & 21.66 & $ 18.45 \pm 0.09 $ & 0.42 & 0.00\cr
ubi1-9 & 01 10 02.19 & -02 28 44.0 & 20.83 & $ 18.50 \pm 0.10 $ & 0.48 & 0.00\cr
ubi1-10 & 01 10 00.98 & -02 28 44.9 & 20.64 & $ 18.18 \pm 0.08 $ & 0.50 & 0.03\cr
ubi1-11 & 01 10 00.95 & -02 28 18.1 & 21.54 & $ 18.82 \pm 0.12 $ & 0.35 & 0.06\cr
ubi1-12 & 01 10 00.73 & -02 28 18.6 & 22.54 & $ 25.14 \pm 43.41 $ & --- & ---\cr
ubi1-13 & 01 10 01.42 & -02 27 39.8 & 22.16 & $ 19.59 \pm 0.35 $ & --- & ---\cr
ubi1-14 & 01 10 03.21 & -02 28 05.9 & 22.73 & $ 19.83 \pm 0.32 $ & --- & ---\cr
ubi1-15 & 01 10 02.69 & -02 28 09.8 & 22.62 & \hbox{No signal} & --- & ---\cr
ubi1-16 & 01 09 59.68 & -02 28 00.4 & 22.58 & $ 19.07 \pm 0.15 $ & --- & ---\cr
ubi1-17 & 01 10 02.23 & -02 28 27.3 & 22.27 & $ 20.01 \pm 0.36 $ & --- & ---\cr
ubi1-18 & 01 09 59.94 & -02 28 15.0 & 21.60 & $ 18.44 \pm 0.14 $ & --- & ---\cr
ubi1-19 & 01 10 01.50 & -02 28 28.1 & 22.69 & $ 22.14 \pm 2.62 $ & --- & ---\cr
ubi1-20 & 01 10 00.98 & -02 28 30.4 & 23.48 & $ 22.05 \pm 2.56 $ & --- & ---\cr
ubi1-21 & 01 10 01.21 & -02 28 32.1 & 23.97 & $ 20.93 \pm 0.87 $ & --- & ---\cr
ubi1-22 & 01 10 00.84 & -02 28 14.2 & 24.65 & \hbox{No signal} & --- & ---\cr
ubi1-23 & 01 10 01.07 & -02 28 28.1 & 22.97 & \hbox{No signal} & --- & ---\cr
ubi1-24 & 01 09 58.24 & -02 28 05.0 & 20.29 & $ 18.81 \pm 0.15 $ & 0.32 & 0.06\cr
ubi1-25 & 01 09 58.18 & -02 27 56.3 & 22.83 & $ 20.04 \pm 0.38 $ & --- & ---\cr
ubi1-26 & 01 09 58.41 & -02 27 45.5 & 22.38 & $ 20.20 \pm 0.48 $ & --- & ---\cr
ubi1-27 & 01 09 57.38 & -02 28 06.2 & 20.18 & $ 18.33 \pm 0.10 $ & 0.26 & 0.05\cr
ubi1-28 & 01 09 58.07 & -02 27 40.0 & 20.12 & $ 17.98 \pm 0.06 $ & 0.33 & 0.10\cr
ubi1-29 & 01 09 58.47 & -02 27 23.8 & 19.67 & $ 17.44 \pm 0.04 $ & 0.29 & 0.11\cr
ubi1-30 & 01 09 58.31 & -02 27 27.5 & 21.59 & $ 20.75 \pm 0.79 $ & 0.48 & 0.00\cr
ubi1-31 & 01 09 58.20 & -02 27 18.7 & 20.24 & $ 18.05 \pm 0.07 $ & --- & ---\cr
ubi1-32 & 01 09 57.25 & -02 27 34.9 & 18.92 & $ 16.45 \pm 0.01 $ & 0.35 & 0.27\cr
ubi1-33 & 01 09 56.24 & -02 27 51.2 & 22.57 & $ 22.70 \pm 4.41 $ & --- & ---\cr
ubi1-34 & 01 09 56.02 & -02 27 17.9 & 22.24 & $ 20.18 \pm 0.42 $ & --- & ---\cr
ubi1-35 & 01 09 55.68 & -02 27 14.2 & 20.97 & $ 19.54 \pm 0.24 $ & 0.28 & 0.07\cr
ubi1-36 & 01 09 58.57 & -02 27 56.5 & 22.60 & $ 21.07 \pm 1.00 $ & --- & ---\cr
ubi1-37 & 01 09 58.98 & -02 27 42.0 & 22.13 & \hbox{No signal} & --- & ---\cr
ubi1-38 & 01 09 59.15 & -02 27 26.9 & 22.55 & $ 20.43 \pm 0.62 $ & --- & ---\cr
ubi1-39 & 01 09 57.95 & -02 27 55.2 & 22.07 & $ 23.59 \pm 9.95 $ & --- & ---\cr
ubi1-40 & 01 09 57.76 & -02 27 56.7 & 23.55 & \hbox{No signal} & --- & ---\cr
ubi1-41 & 01 09 58.51 & -02 27 37.2 & 22.44 & \hbox{No signal} & --- & ---\cr
ubi1-42 & 01 09 58.33 & -02 27 38.4 & 22.68 & \hbox{No signal} & --- & ---\cr
ubi1-43 & 01 09 56.88 & -02 27 32.0 & 21.97 & $ 18.86 \pm 0.14 $ & 0.36 & 0.01\cr
ubi1-44 & 01 09 57.61 & -02 27 02.8 & 21.60 & $ 19.97 \pm 0.39 $ & 0.25 & 0.11\cr
ubi1-45 & 01 09 57.08 & -02 26 56.9 & 21.37 & $ 19.06 \pm 0.20 $ & 0.24 & 0.00\cr
ubi1-46 & 01 09 58.74 & -02 27 42.3 & 22.59 & $ 20.05 \pm 0.45 $ & --- & ---\cr
ubi1-47 & 01 09 55.41 & -02 27 28.0 & 22.86 & \hbox{No signal} & --- & ---\cr
ubi1-48 & 01 09 59.95 & -02 27 04.7 & 20.47 & $ 17.76 \pm 0.05 $ & 0.46 & 0.09\cr
ubi1-49 & 01 09 57.70 & -02 26 23.1 & 21.86 & $ 19.49 \pm 0.32 $ & 0.29 & 0.02\cr
ubi1-50 & 01 09 58.86 & -02 26 23.9 & 19.57 & $ 17.23 \pm 0.03 $ & 0.62 & 0.03\cr
ubi1-51 & 01 09 58.50 & -02 26 18.7 & 21.16 & $ 19.05 \pm 0.16 $ & 0.27 & 0.14\cr
ubi1-52 & 01 10 00.25 & -02 26 23.6 & 22.07 & $ 20.86 \pm 0.76 $ & --- & ---\cr
ubi1-53 & 01 09 59.78 & -02 26 15.1 & 22.09 & \hbox{No signal} & --- & ---\cr
ubi1-54 & 01 10 00.96 & -02 26 22.8 & 22.12 & $ 21.76 \pm 1.90 $ & --- & ---\cr
ubi1-55 & 01 09 58.87 & -02 25 56.7 & 20.45 & $ 18.48 \pm 0.12 $ & 0.23 & 0.05\cr
ubi1-56 & 01 10 01.04 & -02 26 13.4 & 20.43 & $ 17.41 \pm 0.03 $ & 0.38 & 0.10\cr
ubi1-58 & 01 09 59.60 & -02 26 57.8 & 21.94 & $ 20.74 \pm 0.81 $ & 0.16 & 0.05\cr
ubi1-59 & 01 10 00.57 & -02 27 01.1 & 22.17 & $ 19.88 \pm 0.38 $ & --- & ---\cr
ubi1-60 & 01 10 00.42 & -02 26 54.3 & 22.09 & $ 19.79 \pm 0.32 $ & --- & ---\cr
ubi1-61 & 01 10 00.50 & -02 26 49.7 & 21.89 & $ 20.00 \pm 0.36 $ & 0.16 & 0.00\cr
ubi1-62 & 01 09 58.66 & -02 26 25.1 & 20.69 & $ 18.36 \pm 0.09 $ & 0.25 & 0.00\cr
ubi1-63 & 01 10 01.59 & -02 26 39.9 & 21.64 & $ 21.05 \pm 0.93 $ & 0.28 & 0.02\cr
ubi1-64 & 01 10 01.03 & -02 26 29.7 & 20.89 & $ 19.36 \pm 0.20 $ & --- & ---\cr
ubi1-65 & 01 09 59.49 & -02 26 16.6 & 21.54 & $ 18.78 \pm 0.12 $ & 0.47 & 0.02\cr
ubi1-66 & 01 10 01.65 & -02 26 28.6 & 22.63 & $ 19.87 \pm 0.32 $ & --- & ---\cr
ubi1-67 & 01 10 02.29 & -02 26 16.7 & 20.74 & $ 17.47 \pm 0.04 $ & 0.23 & 0.07\cr
ubi1-68 & 01 10 01.16 & -02 26 05.6 & 21.60 & $ 19.64 \pm 0.30 $ & 0.26 & 0.08\cr
ubi1-69 & 01 10 00.96 & -02 26 30.1 & 20.95 & $ 19.80 \pm 0.30 $ & 0.26 & 0.08\cr
ubi1-71 & 01 10 00.34 & -02 26 11.1 & 25.93 & $ 20.45 \pm 0.54 $ & --- & ---\cr
ueh0-2 & 00 53 24.41 & 12 33 50.3 & 20.58 & $ 17.88 \pm 0.07 $ & 0.28 & 0.09\cr
ueh0-7 & 00 53 24.12 & 12 33 01.3 & 22.86 & $ 20.11 \pm 0.44 $ & --- & ---\cr
ueh0-9 & 00 53 24.83 & 12 32 51.3 & 22.86 & \hbox{No signal} & --- & ---\cr
ueh0-11 & 00 53 24.70 & 12 32 58.5 & 22.75 & $ 19.34 \pm 0.31 $ & --- & ---\cr
ueh0-13 & 00 53 24.03 & 12 33 29.1 & 22.88 & $ 21.03 \pm 0.88 $ & --- & ---\cr
ueh0-16 & 00 53 23.66 & 12 33 15.3 & 22.30 & $ 19.23 \pm 0.15 $ & --- & ---\cr
ueh0-18 & 00 53 24.63 & 12 33 05.6 & 23.06 & \hbox{No signal} & --- & ---\cr
ueh0-23 & 00 53 24.68 & 12 33 06.4 & 22.88 & $ 22.36 \pm 3.91 $ & --- & ---\cr
ueh0-24 & 00 53 24.56 & 12 33 01.7 & 99.99 & \hbox{No signal} & --- & ---\cr
ueh0-27 & 00 53 23.03 & 12 33 32.1 & 20.75 & $ 17.59 \pm 0.03 $ & 0.54 & 0.02\cr
ueh0-28 & 00 53 23.05 & 12 33 53.9 & 20.94 & $ 18.49 \pm 0.09 $ & 0.46 & 0.09\cr
ueh0-29 & 00 53 21.82 & 12 33 07.4 & 20.71 & $ 17.74 \pm 0.04 $ & 0.49 & 0.08\cr
ueh0-30 & 00 53 21.73 & 12 33 23.6 & 20.71 & $ 17.77 \pm 0.04 $ & 0.39 & 0.12\cr
ueh0-31 & 00 53 21.28 & 12 33 23.1 & 22.51 & $ 19.84 \pm 0.25 $ & --- & ---\cr
ueh0-32 & 00 53 21.96 & 12 33 57.8 & 22.18 & $ 19.62 \pm 0.21 $ & --- & ---\cr
ueh0-33 & 00 53 20.73 & 12 33 07.0 & 21.26 & $ 18.11 \pm 0.05 $ & 0.33 & 0.06\cr
ueh0-34 & 00 53 21.17 & 12 33 50.1 & 21.61 & $ 18.72 \pm 0.09 $ & 0.42 & 0.00\cr
ueh0-35 & 00 53 20.45 & 12 33 20.2 & 21.37 & $ 18.63 \pm 0.08 $ & 0.25 & 0.14\cr
ueh0-36 & 00 53 19.83 & 12 33 11.6 & 21.35 & $ 18.05 \pm 0.06 $ & 0.48 & 0.03\cr
ueh0-37 & 00 53 19.43 & 12 33 44.8 & 22.50 & $ 21.26 \pm 1.26 $ & --- & ---\cr
ueh0-38 & 00 53 21.74 & 12 33 15.7 & 23.62 & $ 20.67 \pm 0.54 $ & --- & ---\cr
ueh0-39 & 00 53 20.81 & 12 33 21.9 & 22.93 & $ 19.77 \pm 0.24 $ & --- & ---\cr
ueh0-40 & 00 53 20.40 & 12 33 44.4 & 22.19 & $ 18.82 \pm 0.09 $ & --- & ---\cr
ueh0-41 & 00 53 20.51 & 12 33 22.8 & 22.01 & $ 18.95 \pm 0.11 $ & --- & ---\cr
ueh0-42 & 00 53 20.25 & 12 34 31.3 & 21.25 & $ 18.57 \pm 0.09 $ & 0.21 & 0.08\cr
ueh0-43 & 00 53 23.68 & 12 34 17.0 & 21.57 & $ 18.83 \pm 0.11 $ & 0.30 & 0.05\cr
ueh0-44 & 00 53 22.82 & 12 34 25.1 & 22.64 & $ 19.88 \pm 0.25 $ & --- & ---\cr
ueh0-45 & 00 53 21.41 & 12 34 38.1 & 22.93 & $ 19.52 \pm 0.19 $ & --- & ---\cr
ueh0-46 & 00 53 23.32 & 12 34 33.6 & 21.29 & $ 18.91 \pm 0.10 $ & 0.27 & 0.05\cr
ueh0-47 & 00 53 22.86 & 12 34 41.2 & 22.42 & $ 20.17 \pm 0.32 $ & --- & ---\cr
ueh0-48 & 00 53 20.68 & 12 34 54.7 & 22.29 & $ 20.72 \pm 0.58 $ & --- & ---\cr
ueh0-49 & 00 53 23.57 & 12 34 45.3 & 20.75 & $ 17.85 \pm 0.04 $ & 0.32 & 0.05\cr
ueh0-50 & 00 53 20.82 & 12 35 11.3 & 21.61 & $ 19.52 \pm 0.21 $ & 0.20 & 0.09\cr
ueh0-51 & 00 53 22.42 & 12 35 05.3 & 21.91 & $ 19.30 \pm 0.15 $ & 0.36 & 0.00\cr
ueh0-52 & 00 53 23.79 & 12 35 00.1 & 23.04 & $ 19.82 \pm 0.27 $ & --- & ---\cr
ueh0-53 & 00 53 22.48 & 12 35 13.1 & 21.87 & $ 19.52 \pm 0.22 $ & 0.23 & 0.01\cr
ueh0-54 & 00 53 21.54 & 12 34 29.8 & 21.96 & $ 18.67 \pm 0.09 $ & --- & ---\cr
ueh0-55 & 00 53 21.18 & 12 35 17.9 & 21.81 & $ 18.21 \pm 0.08 $ & 0.31 & 0.17\cr
ueh0-56 & 00 53 20.98 & 12 34 34.2 & 23.08 & $ 20.55 \pm 0.47 $ & --- & ---\cr
ueh0-57 & 00 53 23.59 & 12 34 53.7 & 22.35 & \hbox{No signal} & --- & ---\cr
uim0-1 & 03 55 33.45 & 09 43 01.8 & 20.40 & $ 17.30 \pm 0.02 $ & 0.50 & 0.04\cr
uim0-2 & 03 55 33.60 & 09 42 46.9 & 22.10 & $ 19.51 \pm 0.18 $ & --- & ---\cr
uim0-3 & 03 55 31.99 & 09 43 04.7 & 22.06 & $ 18.43 \pm 0.07 $ & --- & ---\cr
uim0-4 & 03 55 32.56 & 09 42 53.7 & 22.89 & \hbox{No signal} & --- & ---\cr
uim0-5 & 03 55 31.72 & 09 42 57.2 & 21.93 & $ 19.28 \pm 0.16 $ & --- & ---\cr
uim0-6 & 03 55 32.34 & 09 42 46.9 & 22.09 & $ 19.99 \pm 0.27 $ & --- & ---\cr
uim0-7 & 03 55 33.07 & 09 42 32.8 & 21.56 & $ 18.36 \pm 0.06 $ & 0.48 & 0.03\cr
uim0-8 & 03 55 31.78 & 09 42 45.9 & 20.14 & $ 17.21 \pm 0.02 $ & 0.48 & 0.02\cr
uim0-9 & 03 55 31.13 & 09 42 41.1 & 21.51 & $ 18.35 \pm 0.06 $ & 0.46 & 0.00\cr
uim0-10 & 03 55 31.34 & 09 42 29.0 & 21.55 & $ 18.63 \pm 0.08 $ & 0.19 & 0.06\cr
uim0-11 & 03 55 31.50 & 09 42 15.1 & 21.23 & $ 19.29 \pm 0.15 $ & 0.23 & 0.32\cr
uim0-12 & 03 55 33.29 & 09 43 08.0 & 22.40 & $ 19.24 \pm 0.18 $ & --- & ---\cr
uim0-13 & 03 55 34.31 & 09 42 23.6 & 22.11 & $ 19.86 \pm 0.34 $ & --- & ---\cr
uim0-14 & 03 55 34.06 & 09 42 18.8 & 22.79 & $ 19.27 \pm 0.20 $ & --- & ---\cr
uim0-15 & 03 55 33.74 & 09 42 18.8 & 21.76 & $ 17.80 \pm 0.04 $ & --- & ---\cr
uim0-16 & 03 55 32.44 & 09 42 39.4 & 22.96 & \hbox{No signal} & --- & ---\cr
uim0-17 & 03 55 30.58 & 09 43 23.3 & 22.23 & $ 19.71 \pm 0.26 $ & --- & ---\cr
uim0-18 & 03 55 29.22 & 09 43 27.9 & 20.89 & $ 18.30 \pm 0.06 $ & 0.20 & 0.17\cr
uim0-19 & 03 55 29.86 & 09 43 43.5 & 19.80 & $ 16.87 \pm 0.02 $ & 0.30 & 0.06\cr
uim0-20 & 03 55 29.33 & 09 43 44.8 & 21.98 & $ 20.54 \pm 0.54 $ & --- & ---\cr
uim0-21 & 03 55 28.79 & 09 43 37.0 & 22.11 & $ 20.21 \pm 0.38 $ & --- & ---\cr
uim0-22 & 03 55 28.05 & 09 43 26.2 & 22.19 & $ 19.54 \pm 0.19 $ & --- & ---\cr
uim0-23 & 03 55 28.51 & 09 43 34.5 & 22.57 & $ 20.20 \pm 0.41 $ & --- & ---\cr
uim0-24 & 03 55 28.04 & 09 43 30.6 & 22.00 & $ 20.04 \pm 0.32 $ & --- & ---\cr
uim0-25 & 03 55 29.29 & 09 44 01.0 & 20.33 & $ 17.44 \pm 0.03 $ & 0.50 & 0.04\cr
uim0-26 & 03 55 27.35 & 09 43 29.7 & 21.99 & $ 20.09 \pm 0.30 $ & 0.25 & 0.00\cr
uim0-27 & 03 55 28.01 & 09 43 43.3 & 21.73 & $ 19.62 \pm 0.22 $ & 0.16 & 0.13\cr
uim0-28 & 03 55 28.54 & 09 43 53.8 & 21.73 & $ 19.05 \pm 0.12 $ & 0.49 & 0.02\cr
uim0-29 & 03 55 28.27 & 09 43 40.1 & 20.97 & $ 17.65 \pm 0.04 $ & 0.21 & 0.90\cr
uim0-30 & 03 55 28.49 & 09 43 40.6 & 21.29 & $ 17.63 \pm 0.04 $ & 0.36 & 0.06\cr
uim0-31 & 03 55 27.98 & 09 43 34.8 & 22.07 & $ 19.11 \pm 0.13 $ & --- & ---\cr
uim0-32 & 03 55 29.05 & 09 43 46.7 & 23.32 & \hbox{No signal} & --- & ---\cr
uim0-33 & 03 55 30.66 & 09 44 18.5 & 21.83 & \hbox{No signal} & 0.20 & 0.08\cr
uim0-34 & 03 55 31.05 & 09 44 14.0 & 21.76 & $ 18.77 \pm 0.12 $ & 0.22 & 0.49\cr
uim0-35 & 03 55 32.01 & 09 44 03.9 & 22.44 & $ 20.88 \pm 0.67 $ & --- & ---\cr
uim0-36 & 03 55 31.12 & 09 44 17.7 & 22.39 & $ 19.31 \pm 0.18 $ & --- & ---\cr
uim0-37 & 03 55 31.53 & 09 44 12.9 & 21.48 & $ 17.99 \pm 0.05 $ & 0.43 & 0.04\cr
uim0-38 & 03 55 31.52 & 09 44 23.5 & 20.79 & $ 18.10 \pm 0.05 $ & 0.31 & 0.17\cr
uim0-39 & 03 55 32.72 & 09 44 27.3 & 23.50 & $ 20.28 \pm 0.38 $ & --- & ---\cr
uim0-40 & 03 55 32.17 & 09 44 45.2 & 22.57 & $ 19.48 \pm 0.17 $ & --- & ---\cr
uim0-41 & 03 55 33.03 & 09 44 44.0 & 18.62 & $ 15.39 \pm 0.01 $ & 0.44 & 0.13\cr
uim0-42 & 03 55 32.00 & 09 44 59.4 & 21.74 & $ 19.70 \pm 0.30 $ & 0.30 & 0.03\cr
uim0-43 & 03 55 34.03 & 09 44 41.4 & 20.59 & $ 17.02 \pm 0.02 $ & 0.53 & 0.05\cr
uim0-44 & 03 55 32.32 & 09 43 52.3 & 21.98 & $ 19.95 \pm 0.37 $ & 0.25 & 0.04\cr
uim0-45 & 03 55 31.59 & 09 44 02.7 & 22.44 & $ 19.14 \pm 0.15 $ & --- & ---\cr
uim0-46 & 03 55 33.03 & 09 43 53.0 & 22.54 & $ 19.35 \pm 0.23 $ & --- & ---\cr
uim0-47 & 03 55 33.14 & 09 44 15.6 & 22.24 & $ 19.20 \pm 0.13 $ & --- & ---\cr
uim0-48 & 03 55 32.43 & 09 44 32.6 & 22.06 & $ 20.11 \pm 0.32 $ & --- & ---\cr
uim0-49 & 03 55 33.52 & 09 44 19.2 & 21.60 & $ 19.41 \pm 0.17 $ & 0.29 & 0.00\cr
uim0-51 & 03 55 32.89 & 09 44 24.4 & 23.22 & $ 21.74 \pm 1.47 $ & --- & ---\cr
uim0-52 & 03 55 32.17 & 09 44 37.3 & 22.58 & $ 20.12 \pm 0.31 $ & --- & ---\cr
umd4-6 & 21 51 06.68 & 29 00 16.5 & 22.97 & \hbox{No signal} & --- & ---\cr
umd4-9 & 21 51 05.61 & 29 00 25.3 & 22.45 & $ 24.11 \pm 29.19 $ & --- & ---\cr
umd4-12 & 21 51 06.01 & 29 00 27.5 & 22.32 & $ 19.63 \pm 0.44 $ & --- & ---\cr
umd4-13 & 21 51 05.86 & 29 00 32.5 & 19.42 & $ 17.29 \pm 0.05 $ & 0.31 & 0.08\cr
umd4-17 & 21 51 05.86 & 29 00 53.2 & 23.51 & \hbox{No signal} & --- & ---\cr
umd4-20 & 21 51 06.06 & 29 01 02.1 & 22.36 & $ 18.39 \pm 0.14 $ & --- & ---\cr
umd4-21 & 21 51 05.81 & 29 01 02.3 & 23.07 & $ 19.22 \pm 0.34 $ & --- & ---\cr
umd4-23 & 21 51 06.22 & 29 01 06.0 & 22.64 & $ 19.83 \pm 0.55 $ & --- & ---\cr
umd4-25 & 21 51 06.68 & 29 01 07.7 & 22.78 & $ 18.62 \pm 0.16 $ & --- & ---\cr
umd4-28 & 21 51 07.40 & 29 00 15.4 & 23.07 & $ 20.19 \pm 0.53 $ & --- & ---\cr
umd4-29 & 21 51 07.54 & 29 01 07.0 & 22.11 & $ 20.38 \pm 0.79 $ & --- & ---\cr
umd4-30 & 21 51 07.57 & 29 00 58.9 & 23.26 & $ 20.27 \pm 0.58 $ & --- & ---\cr
umd4-31 & 21 51 07.82 & 29 00 27.2 & 22.56 & $ 20.78 \pm 0.86 $ & --- & ---\cr
umd4-32 & 21 51 07.80 & 29 00 51.7 & 23.29 & \hbox{No signal} & --- & ---\cr
umd4-33 & 21 51 07.88 & 29 01 01.8 & 23.36 & $ 20.52 \pm 0.74 $ & --- & ---\cr
umd4-34 & 21 51 08.12 & 29 00 18.5 & 23.47 & $ 21.95 \pm 2.62 $ & --- & ---\cr
umd4-35 & 21 51 08.33 & 29 00 40.4 & 23.08 & $ 20.27 \pm 0.53 $ & --- & ---\cr
umd4-36 & 21 51 08.31 & 29 00 50.8 & 23.67 & $ 19.49 \pm 0.26 $ & --- & ---\cr
umd4-37 & 21 51 08.62 & 29 00 00.7 & 24.27 & \hbox{No signal} & --- & ---\cr
umd4-38 & 21 51 08.82 & 29 00 56.8 & 21.26 & $ 18.83 \pm 0.15 $ & 0.42 & 0.08\cr
umd4-39 & 21 51 08.95 & 29 00 19.3 & 21.81 & $ 18.89 \pm 0.15 $ & 0.34 & 0.07\cr
umd4-40 & 21 51 09.15 & 29 00 21.0 & 21.66 & $ 19.13 \pm 0.19 $ & 0.39 & 0.04\cr
umd4-41 & 21 51 09.08 & 29 01 05.2 & 19.92 & $ 17.18 \pm 0.04 $ & 0.30 & 0.15\cr
umd4-42 & 21 51 08.93 & 29 00 01.4 & 23.34 & \hbox{No signal} & --- & ---\cr
umd4-44 & 21 51 09.37 & 29 00 08.6 & 18.97 & $ 16.96 \pm 0.03 $ & 0.34 & 0.11\cr
umd4-45 & 21 51 09.92 & 28 59 56.9 & 23.78 & $ 21.85 \pm 2.28 $ & --- & ---\cr
umd4-46 & 21 51 09.99 & 29 01 01.1 & 23.25 & $ 19.84 \pm 0.40 $ & --- & ---\cr
umd4-47 & 21 51 10.08 & 29 00 17.8 & 23.97 & \hbox{No signal} & --- & ---\cr
umd4-48 & 21 51 10.13 & 29 00 44.7 & 23.01 & $ 19.53 \pm 0.30 $ & --- & ---\cr
umd4-49 & 21 51 10.21 & 29 00 00.5 & 24.81 & \hbox{No signal} & --- & ---\cr
umd4-50 & 21 51 10.22 & 29 00 14.2 & 22.73 & $ 19.78 \pm 0.35 $ & --- & ---\cr
umd4-51 & 21 51 10.52 & 29 00 33.0 & 23.20 & $ 18.51 \pm 0.12 $ & --- & ---\cr
umd4-52 & 21 51 10.60 & 29 00 15.1 & 22.41 & $ 19.58 \pm 0.34 $ & --- & ---\cr
umd4-53 & 21 51 11.27 & 29 00 33.8 & 23.30 & \hbox{No signal} & --- & ---\cr
umd4-54 & 21 51 11.65 & 29 00 11.2 & 22.83 & $ 19.08 \pm 0.29 $ & --- & ---\cr
umd4-58 & 21 51 07.22 & 28 59 51.0 & 23.21 & \hbox{No signal} & --- & ---\cr
umd4-59 & 21 51 09.83 & 28 59 48.4 & 20.31 & $ 17.82 \pm 0.06 $ & 0.41 & 0.04\cr
umd4-61 & 21 51 10.26 & 28 59 43.7 & 23.21 & \hbox{No signal} & --- & ---\cr
umd4-62 & 21 51 09.10 & 28 59 38.5 & 20.70 & $ 18.08 \pm 0.07 $ & 0.34 & 0.12\cr
umd4-63 & 21 51 09.79 & 28 59 39.4 & 23.89 & \hbox{No signal} & --- & ---\cr
umd4-64 & 21 51 08.70 & 28 59 30.5 & 23.21 & $ 19.02 \pm 0.17 $ & --- & ---\cr
umd4-65 & 21 51 08.86 & 28 59 28.2 & 20.98 & $ 18.18 \pm 0.08 $ & 0.36 & 0.05\cr
umd4-66 & 21 51 07.61 & 28 59 28.7 & 23.11 & \hbox{No signal} & --- & ---\cr
umd4-68 & 21 51 07.19 & 28 59 22.2 & 22.74 & $ 19.37 \pm 0.23 $ & --- & ---\cr
umd4-69 & 21 51 09.64 & 28 59 18.7 & 23.61 & \hbox{No signal} & --- & ---\cr
umd4-70 & 21 51 10.71 & 28 59 18.8 & 23.97 & $ 20.14 \pm 0.57 $ & --- & ---\cr
umd4-71 & 21 51 07.89 & 28 59 16.7 & 22.70 & $ 19.46 \pm 0.24 $ & --- & ---\cr
umd4-72 & 21 51 10.98 & 28 59 15.6 & 21.84 & $ 19.78 \pm 0.45 $ & --- & ---\cr
umd4-74 & 21 51 09.93 & 28 59 13.7 & 21.38 & $ 19.53 \pm 0.26 $ & 0.28 & 0.12\cr
umd4-75 & 21 51 07.36 & 28 59 12.3 & 23.15 & $ 21.48 \pm 1.60 $ & --- & ---\cr
umd4-76 & 21 51 07.34 & 28 59 07.1 & 21.27 & $ 19.88 \pm 0.37 $ & --- & ---\cr
umd4-77 & 21 51 09.18 & 28 59 07.5 & 23.63 & $ 21.91 \pm 2.45 $ & --- & ---\cr
umd4-78 & 21 51 08.16 & 28 59 05.4 & 23.30 & $ 20.09 \pm 0.44 $ & --- & ---\cr
umd4-82 & 21 51 07.37 & 28 58 48.4 & 23.33 & \hbox{No signal} & --- & ---\cr
uo50-3 & 17 55 26.37 & 18 17 57.4 & 22.23 & $ 20.09 \pm 0.67 $ & --- & ---\cr
uo50-8 & 17 55 25.64 & 18 18 03.8 & 22.98 & $ 21.22 \pm 1.12 $ & --- & ---\cr
uo50-15 & 17 55 24.60 & 18 17 56.2 & 21.06 & $ 19.02 \pm 0.22 $ & 0.22 & 0.31\cr
uo50-19 & 17 55 24.16 & 18 17 57.7 & 20.86 & $ 18.39 \pm 0.10 $ & 0.23 & 0.09\cr
uo50-20 & 17 55 24.05 & 18 17 56.1 & 21.04 & $ 18.43 \pm 0.12 $ & 0.24 & 0.08\cr
uo50-28 & 17 55 23.36 & 18 17 53.6 & 21.93 & $ 22.60 \pm 6.13 $ & 0.30 & 0.31\cr
uo50-34 & 17 55 23.74 & 18 18 06.5 & 22.14 & $ 19.06 \pm 0.14 $ & --- & ---\cr
uo50-35 & 17 55 23.21 & 18 18 05.3 & 23.40 & $ 19.62 \pm 0.25 $ & --- & ---\cr
uo50-36 & 17 55 25.22 & 18 18 15.0 & 22.05 & $ 19.52 \pm 0.18 $ & --- & ---\cr
uo50-37 & 17 55 24.69 & 18 18 13.0 & 22.58 & \hbox{No signal} & --- & ---\cr
uo50-38 & 17 55 25.88 & 18 18 19.1 & 22.81 & $ 20.05 \pm 0.30 $ & --- & ---\cr
uo50-39 & 17 55 26.53 & 18 18 21.6 & 20.93 & $ 18.95 \pm 0.16 $ & --- & ---\cr
uo50-40 & 17 55 25.60 & 18 18 24.1 & 20.79 & $ 18.39 \pm 0.07 $ & --- & ---\cr
uo50-41 & 17 55 23.97 & 18 18 14.4 & 22.02 & $ 21.77 \pm 1.34 $ & --- & ---\cr
uo50-42 & 17 55 24.88 & 18 18 21.7 & 22.80 & $ 20.55 \pm 0.45 $ & --- & ---\cr
uo50-43 & 17 55 22.00 & 18 18 10.6 & 22.68 & \hbox{No signal} & --- & ---\cr
uo50-44 & 17 55 25.40 & 18 18 28.6 & 21.37 & $ 18.68 \pm 0.09 $ & --- & ---\cr
uo50-45 & 17 55 25.61 & 18 18 33.4 & 21.16 & $ 17.67 \pm 0.04 $ & 0.39 & 0.14\cr
uo50-46 & 17 55 25.55 & 18 18 30.8 & 21.37 & $ 17.79 \pm 0.04 $ & --- & ---\cr
uo50-47 & 17 55 25.47 & 18 18 27.4 & 21.56 & $ 18.79 \pm 0.11 $ & --- & ---\cr
uo50-48 & 17 55 24.21 & 18 18 26.1 & 23.55 & $ 19.99 \pm 0.29 $ & --- & ---\cr
uo50-49 & 17 55 23.31 & 18 18 22.6 & 22.22 & $ 20.29 \pm 0.35 $ & --- & ---\cr
uo50-50 & 17 55 25.19 & 18 18 31.2 & 23.61 & \hbox{No signal} & --- & ---\cr
uo50-51 & 17 55 24.13 & 18 18 29.5 & 23.05 & $ 21.11 \pm 0.83 $ & --- & ---\cr
uo50-52 & 17 55 22.04 & 18 18 21.4 & 23.51 & $ 19.83 \pm 0.30 $ & --- & ---\cr
uo50-53 & 17 55 25.60 & 18 18 41.2 & 21.14 & $ 18.78 \pm 0.11 $ & --- & ---\cr
uo50-54 & 17 55 25.76 & 18 18 38.4 & 20.81 & $ 18.33 \pm 0.07 $ & --- & ---\cr
uo50-55 & 17 55 25.52 & 18 18 38.3 & 21.67 & $ 20.20 \pm 0.40 $ & --- & ---\cr
uo50-56 & 17 55 24.39 & 18 18 35.3 & 24.33 & \hbox{No signal} & --- & ---\cr
uo50-57 & 17 55 24.11 & 18 18 38.2 & 23.14 & $ 19.68 \pm 0.23 $ & --- & ---\cr
uo50-58 & 17 55 23.92 & 18 18 40.4 & 20.33 & $ 17.75 \pm 0.04 $ & 0.24 & 0.35\cr
uo50-59 & 17 55 21.65 & 18 18 31.5 & 22.76 & $ 19.55 \pm 0.30 $ & --- & ---\cr
uo50-60 & 17 55 21.78 & 18 18 33.8 & 21.59 & \hbox{No signal} & --- & ---\cr
uo50-61 & 17 55 24.30 & 18 18 48.9 & 18.34 & $ 17.09 \pm 0.02 $ & --- & ---\cr
uo50-62 & 17 55 23.70 & 18 18 45.7 & 19.78 & $ 17.33 \pm 0.03 $ & 0.67 & 0.36\cr
uo50-63 & 17 55 25.94 & 18 18 57.2 & 22.95 & $ 20.50 \pm 0.50 $ & --- & ---\cr
uo50-64 & 17 55 23.76 & 18 18 56.0 & 20.85 & $ 18.34 \pm 0.07 $ & --- & ---\cr
uo50-65 & 17 55 23.58 & 18 18 55.3 & 21.59 & $ 20.37 \pm 0.45 $ & --- & ---\cr
uo50-66 & 17 55 24.45 & 18 18 55.9 & 21.47 & $ 18.73 \pm 0.11 $ & --- & ---\cr
uo50-67 & 17 55 22.41 & 18 18 47.3 & 22.23 & $ 19.13 \pm 0.14 $ & --- & ---\cr
uo50-68 & 17 55 23.01 & 18 18 52.9 & 23.27 & $ 20.13 \pm 0.30 $ & --- & ---\cr
uo50-70 & 17 55 24.89 & 18 19 06.8 & 23.53 & $ 20.05 \pm 0.39 $ & --- & ---\cr
uo50-71 & 17 55 23.78 & 18 19 02.6 & 22.29 & $ 20.42 \pm 0.46 $ & --- & ---\cr
uo50-72 & 17 55 22.14 & 18 18 57.9 & 21.46 & $ 18.78 \pm 0.11 $ & --- & ---\cr
uo50-73 & 17 55 21.44 & 18 18 54.4 & 22.43 & \hbox{No signal} & --- & ---\cr
uo50-74 & 17 55 21.48 & 18 19 00.8 & 22.35 & $ 19.43 \pm 0.25 $ & --- & ---\cr
uo50-76 & 17 55 23.86 & 18 19 12.2 & 21.67 & $ 21.90 \pm 2.50 $ & 0.38 & -0.06\cr
uo50-77 & 17 55 22.83 & 18 19 07.8 & 22.26 & \hbox{No signal} & --- & ---\cr
uo50-78 & 17 55 21.29 & 18 19 01.8 & 23.14 & $ 20.16 \pm 0.58 $ & --- & ---\cr
uo50-80 & 17 55 27.19 & 18 18 40.5 & 21.73 & $ 20.16 \pm 0.66 $ & 0.71 & 0.17\cr
uo50-81 & 17 55 26.95 & 18 18 54.6 & 21.75 & $ 19.65 \pm 0.31 $ & 0.44 & 0.01\cr
uo50-83 & 17 55 27.08 & 18 19 01.6 & 20.98 & $ 17.53 \pm 0.05 $ & 0.48 & -0.08\cr
uop0-2 & 07 50 47.31 & 14 40 21.2 & 22.36 & $ 19.41 \pm 0.22 $ & --- & ---\cr
uop0-4 & 07 50 47.18 & 14 40 16.1 & 21.54 & $ 19.31 \pm 0.22 $ & 0.28 & -0.07\cr
uop0-10 & 07 50 45.34 & 14 40 08.1 & 24.75 & $ 20.83 \pm 0.98 $ & --- & ---\cr
uop0-11 & 07 50 44.71 & 14 40 04.0 & 22.80 & $ 21.71 \pm 2.30 $ & --- & ---\cr
uop0-14 & 07 50 43.94 & 14 40 01.7 & 99.99 & $ 21.62 \pm 2.18 $ & --- & ---\cr
uop0-16 & 07 50 43.62 & 14 40 06.1 & 24.23 & \hbox{No signal} & --- & ---\cr
uop0-17 & 07 50 45.96 & 14 40 26.8 & 22.83 & $ 20.53 \pm 0.50 $ & --- & ---\cr
uop0-18 & 07 50 46.54 & 14 40 32.5 & 21.87 & $ 18.86 \pm 0.12 $ & 0.29 & 0.00\cr
uop0-19 & 07 50 46.57 & 14 40 34.0 & 21.81 & $ 18.83 \pm 0.13 $ & --- & ---\cr
uop0-20 & 07 50 44.16 & 14 40 15.3 & 22.54 & $ 19.53 \pm 0.24 $ & --- & ---\cr
uop0-21 & 07 50 46.82 & 14 40 38.0 & 23.73 & $ 20.18 \pm 0.41 $ & --- & ---\cr
uop0-22 & 07 50 43.95 & 14 40 15.7 & 21.34 & $ 20.45 \pm 0.53 $ & 0.25 & -0.11\cr
uop0-23 & 07 50 43.29 & 14 40 10.5 & 99.99 & $ 20.54 \pm 0.81 $ & --- & ---\cr
uop0-24 & 07 50 46.83 & 14 40 45.9 & 21.47 & $ 18.42 \pm 0.07 $ & 0.10 & 0.03\cr
uop0-25 & 07 50 46.43 & 14 40 49.6 & 22.73 & $ 22.67 \pm 3.62 $ & --- & ---\cr
uop0-26 & 07 50 43.01 & 14 40 23.0 & 22.81 & $ 19.01 \pm 0.17 $ & --- & ---\cr
uop0-27 & 07 50 46.81 & 14 40 56.2 & 22.23 & $ 20.14 \pm 0.33 $ & --- & ---\cr
uop0-28 & 07 50 42.73 & 14 40 29.6 & 21.54 & $ 19.70 \pm 0.33 $ & --- & ---\cr
uop0-29 & 07 50 46.61 & 14 41 04.3 & 22.50 & $ 19.93 \pm 0.26 $ & --- & ---\cr
uop0-30 & 07 50 43.06 & 14 40 37.8 & 21.09 & $ 18.92 \pm 0.14 $ & 0.24 & 0.09\cr
uop0-31 & 07 50 44.83 & 14 40 28.1 & 22.47 & $ 19.50 \pm 0.24 $ & --- & ---\cr
uop0-32 & 07 50 43.94 & 14 40 43.1 & 16.71 & $ 14.68 \pm 0.00 $ & 0.60 & 0.00\cr
uop0-33 & 07 50 43.64 & 14 40 38.7 & 22.86 & $ 18.62 \pm 0.14 $ & --- & ---\cr
uop0-34 & 07 50 44.89 & 14 40 31.8 & 19.88 & $ 16.89 \pm 0.03 $ & --- & ---\cr
uop0-35 & 07 50 45.47 & 14 41 03.0 & 22.86 & $ 20.84 \pm 0.67 $ & --- & ---\cr
uop0-36 & 07 50 45.14 & 14 41 16.9 & 21.75 & $ 18.20 \pm 0.10 $ & --- & ---\cr
uop0-37 & 07 50 44.88 & 14 41 11.0 & 17.64 & $ 15.62 \pm 0.01 $ & 0.56 & 0.03\cr
uop0-41 & 07 50 47.54 & 14 40 55.9 & 21.22 & $ 16.08 \pm 0.01 $ & --- & ---\cr
uop0-42 & 07 50 46.78 & 14 41 17.9 & 22.12 & $ 17.31 \pm 0.03 $ & --- & ---\cr
uop0-43 & 07 50 46.77 & 14 41 15.8 & 22.52 & $ 16.69 \pm 0.01 $ & --- & ---\cr
uop0-44 & 07 50 47.11 & 14 41 18.6 & 21.75 & $ 19.12 \pm 0.15 $ & 0.35 & -0.01\cr
uop0-45 & 07 50 47.80 & 14 41 21.4 & 22.24 & $ 19.07 \pm 0.16 $ & --- & ---\cr
uop0-46 & 07 50 47.14 & 14 41 38.0 & 21.32 & $ 19.02 \pm 0.16 $ & 0.43 & -0.10\cr
uop0-47 & 07 50 47.39 & 14 41 37.4 & 22.51 & $ 18.97 \pm 0.14 $ & --- & ---\cr
uop0-48 & 07 50 49.10 & 14 41 04.5 & 22.01 & $ 18.36 \pm 0.07 $ & --- & ---\cr
uop0-49 & 07 50 48.48 & 14 41 22.8 & 23.48 & $ 20.16 \pm 0.37 $ & --- & ---\cr
uop0-50 & 07 50 47.76 & 14 41 47.3 & 23.12 & $ 20.00 \pm 0.42 $ & --- & ---\cr
uop0-51 & 07 50 48.96 & 14 41 33.8 & 21.64 & $ 17.54 \pm 0.04 $ & --- & ---\cr
uop0-52 & 07 50 49.59 & 14 41 17.2 & 23.19 & $ 21.47 \pm 1.27 $ & --- & ---\cr
uop0-53 & 07 50 49.77 & 14 41 14.3 & 22.05 & $ 19.60 \pm 0.23 $ & --- & ---\cr
uop0-54 & 07 50 50.26 & 14 41 13.0 & 21.14 & $ 17.71 \pm 0.04 $ & 0.26 & -0.01\cr
uop0-55 & 07 50 50.08 & 14 41 23.4 & 23.01 & \hbox{No signal} & --- & ---\cr
uop0-56 & 07 50 49.21 & 14 41 53.9 & 22.86 & $ 20.06 \pm 0.50 $ & --- & ---\cr
uop0-57 & 07 50 51.29 & 14 41 20.1 & 20.69 & $ 18.45 \pm 0.09 $ & 0.37 & -0.01\cr
uop0-58 & 07 50 51.20 & 14 41 32.3 & 22.38 & $ 19.23 \pm 0.19 $ & --- & ---\cr
uop0-60 & 07 50 51.07 & 14 41 37.5 & 22.43 & $ 20.01 \pm 0.37 $ & --- & ---\cr
uop0-61 & 07 50 50.53 & 14 41 51.8 & 22.97 & $ 19.57 \pm 0.32 $ & --- & ---\cr
uop0-62 & 07 50 51.45 & 14 41 42.5 & 22.46 & $ 19.30 \pm 0.27 $ & --- & ---\cr
usa0-47 & 17 12 24.58 & 33 36 26.8 & 23.49 & $ 21.72 \pm 3.11 $ & --- & ---\cr
usa0-56 & 17 12 24.80 & 33 36 41.9 & 23.27 & $ 19.48 \pm 0.47 $ & --- & ---\cr
usa0-57 & 17 12 25.88 & 33 36 36.1 & 21.99 & $ 21.00 \pm 1.23 $ & 0.46 & 0.06\cr
usa0-70 & 17 12 25.18 & 33 36 41.6 & 23.68 & $ 20.25 \pm 0.87 $ & --- & ---\cr
usa0-74 & 17 12 24.64 & 33 35 55.8 & 23.62 & \hbox{No signal} & --- & ---\cr
usa0-76 & 17 12 27.37 & 33 36 25.9 & 22.10 & $ 19.42 \pm 0.19 $ & --- & ---\cr
usa0-77 & 17 12 26.68 & 33 36 07.9 & 22.83 & $ 22.03 \pm 2.12 $ & --- & ---\cr
usa0-78 & 17 12 25.19 & 33 35 47.3 & 19.03 & $ 16.44 \pm 0.02 $ & 0.67 & 0.12\cr
usa0-79 & 17 12 28.40 & 33 36 34.7 & 23.65 & $ 23.71 \pm 12.57 $ & --- & ---\cr
usa0-80 & 17 12 28.04 & 33 36 26.5 & 22.94 & $ 21.16 \pm 0.99 $ & --- & ---\cr
usa0-81 & 17 12 28.40 & 33 36 27.2 & 22.75 & $ 19.70 \pm 0.27 $ & --- & ---\cr
usa0-82 & 17 12 27.06 & 33 35 57.5 & 21.12 & $ 18.53 \pm 0.08 $ & 0.55 & 0.08\cr
usa0-83 & 17 12 26.30 & 33 35 38.2 & 24.13 & $ 20.91 \pm 0.79 $ & --- & ---\cr
usa0-84 & 17 12 29.35 & 33 36 35.5 & 19.41 & $ 16.78 \pm 0.03 $ & --- & ---\cr
usa0-85 & 17 12 29.52 & 33 36 33.0 & 20.86 & $ 17.22 \pm 0.03 $ & 0.29 & 0.37\cr
usa0-86 & 17 12 28.47 & 33 36 19.7 & 23.83 & $ 20.86 \pm 0.70 $ & --- & ---\cr
usa0-87 & 17 12 27.18 & 33 35 48.5 & 20.58 & $ 17.75 \pm 0.04 $ & --- & ---\cr
usa0-88 & 17 12 28.72 & 33 36 16.2 & 23.47 & \hbox{No signal} & --- & ---\cr
usa0-89 & 17 12 29.55 & 33 36 24.9 & 20.88 & $ 18.42 \pm 0.08 $ & 0.24 & 0.10\cr
usa0-91 & 17 12 29.68 & 33 36 19.1 & 22.00 & $ 19.65 \pm 0.26 $ & --- & ---\cr
usa0-92 & 17 12 27.66 & 33 35 30.9 & 22.90 & $ 19.10 \pm 0.17 $ & --- & ---\cr
usa0-93 & 17 12 27.48 & 33 35 30.1 & 20.95 & $ 17.87 \pm 0.05 $ & 0.65 & 0.11\cr
usa0-94 & 17 12 27.35 & 33 35 27.4 & 22.68 & $ 19.51 \pm 0.29 $ & --- & ---\cr
usa0-95 & 17 12 28.88 & 33 35 57.1 & 22.00 & $ 19.23 \pm 0.16 $ & --- & ---\cr
usa0-96 & 17 12 28.23 & 33 35 43.7 & 23.40 & $ 20.70 \pm 0.58 $ & --- & ---\cr
usa0-97 & 17 12 29.73 & 33 36 12.2 & 22.60 & $ 18.87 \pm 0.13 $ & --- & ---\cr
usa0-98 & 17 12 30.34 & 33 36 07.3 & 21.35 & $ 17.89 \pm 0.06 $ & 0.34 & 0.09\cr
usa0-99 & 17 12 28.01 & 33 35 29.4 & 21.32 & $ 19.32 \pm 0.21 $ & 0.40 & 0.10\cr
usa0-100 & 17 12 29.45 & 33 35 56.4 & 22.34 & $ 19.09 \pm 0.14 $ & --- & ---\cr
usa0-101 & 17 12 29.27 & 33 35 45.7 & 22.92 & $ 18.77 \pm 0.11 $ & --- & ---\cr
usa0-102 & 17 12 31.12 & 33 36 16.5 & 21.98 & $ 18.56 \pm 0.13 $ & 0.46 & 0.03\cr
usa0-103 & 17 12 31.32 & 33 36 13.3 & 22.74 & $ 20.31 \pm 0.77 $ & --- & ---\cr
usa0-104 & 17 12 30.84 & 33 36 09.5 & 24.64 & \hbox{No signal} & --- & ---\cr
usa0-105 & 17 12 30.60 & 33 36 04.2 & 23.23 & \hbox{No signal} & --- & ---\cr
usa0-106 & 17 12 28.72 & 33 35 19.4 & 20.49 & $ 18.83 \pm 0.18 $ & 0.27 & 0.13\cr
usa0-107 & 17 12 29.16 & 33 35 27.3 & 23.66 & $ 20.88 \pm 0.97 $ & --- & ---\cr
usa0-108 & 17 12 29.95 & 33 35 41.0 & 21.64 & $ 19.24 \pm 0.16 $ & 0.30 & 0.29\cr
usa0-109 & 17 12 30.90 & 33 36 00.2 & 23.33 & \hbox{No signal} & --- & ---\cr
usa0-110 & 17 12 30.09 & 33 35 36.8 & 22.84 & $ 20.93 \pm 0.92 $ & --- & ---\cr
usa0-111 & 17 12 30.03 & 33 35 33.8 & 23.22 & $ 19.28 \pm 0.19 $ & --- & ---\cr
usa0-112 & 17 12 30.18 & 33 35 34.0 & 23.42 & $ 19.37 \pm 0.21 $ & --- & ---\cr
usa0-114 & 17 12 29.90 & 33 35 29.6 & 23.82 & \hbox{No signal} & --- & ---\cr
usa0-115 & 17 12 31.45 & 33 35 59.6 & 23.82 & $ 20.67 \pm 0.96 $ & --- & ---\cr
ux40-1 & 15 19 39.35 & 23 52 40.2 & 22.14 & $ 19.81 \pm 0.26 $ & --- & ---\cr
ux40-2 & 15 19 39.00 & 23 52 46.7 & 22.18 & \hbox{No signal} & --- & ---\cr
ux40-3 & 15 19 40.62 & 23 52 21.6 & 19.79 & $ 17.39 \pm 0.04 $ & --- & ---\cr
ux40-4 & 15 19 39.52 & 23 52 41.7 & 22.14 & $ 19.37 \pm 0.16 $ & --- & ---\cr
ux40-5 & 15 19 40.39 & 23 52 25.5 & 21.83 & $ 18.20 \pm 0.06 $ & 0.58 & 0.06\cr
ux40-7 & 15 19 38.86 & 23 53 02.1 & 18.91 & $ 16.76 \pm 0.02 $ & 0.43 & 0.16\cr
ux40-8 & 15 19 41.11 & 23 52 22.6 & 18.57 & $ 16.14 \pm 0.01 $ & 0.62 & 0.04\cr
ux40-9 & 15 19 39.34 & 23 52 55.3 & 22.15 & $ 20.07 \pm 0.31 $ & --- & ---\cr
ux40-10 & 15 19 40.47 & 23 52 33.8 & 22.60 & $ 19.91 \pm 0.28 $ & --- & ---\cr
ux40-11 & 15 19 38.61 & 23 53 14.7 & 21.67 & $ 18.55 \pm 0.10 $ & 0.29 & 0.13\cr
ux40-12 & 15 19 40.72 & 23 52 36.2 & 22.64 & $ 19.41 \pm 0.18 $ & --- & ---\cr
ux40-13 & 15 19 40.54 & 23 52 40.7 & 22.17 & $ 19.61 \pm 0.21 $ & --- & ---\cr
ux40-14 & 15 19 38.79 & 23 53 18.0 & 22.29 & \hbox{No signal} & --- & ---\cr
ux40-15 & 15 19 41.00 & 23 52 35.2 & 23.34 & $ 19.77 \pm 0.25 $ & --- & ---\cr
ux40-16 & 15 19 41.15 & 23 52 43.7 & 20.95 & $ 17.95 \pm 0.05 $ & --- & ---\cr
ux40-17 & 15 19 41.39 & 23 52 41.9 & 19.04 & $ 16.11 \pm 0.01 $ & 0.30 & 0.22\cr
ux40-18 & 15 19 39.86 & 23 53 12.2 & 22.37 & $ 20.18 \pm 0.34 $ & --- & ---\cr
ux40-19 & 15 19 41.92 & 23 52 33.3 & 21.99 & $ 18.98 \pm 0.11 $ & --- & ---\cr
ux40-20 & 15 19 41.13 & 23 52 51.5 & 23.36 & $ 20.18 \pm 0.36 $ & --- & ---\cr
ux40-21 & 15 19 40.91 & 23 52 55.1 & 22.58 & $ 20.26 \pm 0.36 $ & --- & ---\cr
ux40-22 & 15 19 42.28 & 23 52 37.2 & 22.96 & $ 22.41 \pm 2.72 $ & --- & ---\cr
ux40-23 & 15 19 41.34 & 23 53 01.1 & 23.74 & \hbox{No signal} & --- & ---\cr
ux40-24 & 15 19 40.31 & 23 53 25.7 & 21.96 & $ 20.26 \pm 0.42 $ & 0.26 & 0.02\cr
ux40-25 & 15 19 40.96 & 23 53 11.7 & 99.99 & $ 19.39 \pm 0.19 $ & --- & ---\cr
ux40-26 & 15 19 41.29 & 23 53 13.0 & 21.29 & $ 18.21 \pm 0.06 $ & --- & ---\cr
ux40-27 & 15 19 42.94 & 23 52 46.2 & 20.05 & $ 17.96 \pm 0.05 $ & 0.28 & 0.17\cr
ux40-28 & 15 19 40.85 & 23 53 33.8 & 18.11 & $ 16.19 \pm 0.01 $ & --- & ---\cr
ux40-29 & 15 19 43.31 & 23 52 48.3 & 20.05 & $ 18.82 \pm 0.12 $ & 0.28 & 0.43\cr
ux40-30 & 15 19 42.62 & 23 53 02.3 & 24.93 & $ 23.33 \pm 6.21 $ & --- & ---\cr
ux40-31 & 15 19 41.29 & 23 53 34.7 & 30.71 & $ 21.76 \pm 2.33 $ & --- & ---\cr
ux40-32 & 15 19 43.81 & 23 52 48.3 & 20.75 & $ 18.58 \pm 0.10 $ & 0.58 & 0.05\cr
ux40-33 & 15 19 42.99 & 23 53 05.1 & 22.83 & $ 24.64 \pm 22.34 $ & --- & ---\cr
ux40-34 & 15 19 42.51 & 23 53 21.0 & 20.72 & $ 18.69 \pm 0.09 $ & 0.50 & 0.01\cr
ux40-35 & 15 19 42.50 & 23 53 18.4 & 20.96 & $ 18.79 \pm 0.10 $ & --- & ---\cr
ux40-36 & 15 19 43.41 & 23 53 09.5 & 24.45 & $ 20.32 \pm 0.48 $ & --- & ---\cr
ux40-37 & 15 19 38.55 & 23 53 08.8 & 21.81 & $ 18.87 \pm 0.14 $ & 0.48 & 0.04\cr
ux40-38 & 15 19 42.48 & 23 52 57.1 & 24.12 & \hbox{No signal} & --- & ---\cr
ux40-39 & 15 19 42.11 & 23 53 14.3 & 22.91 & $ 21.45 \pm 1.02 $ & --- & ---\cr
ux40-40 & 15 19 42.76 & 23 53 06.6 & 22.91 & $ 21.16 \pm 0.84 $ & --- & ---\cr
ux40-41 & 15 19 39.44 & 23 52 34.6 & 22.18 & $ 18.43 \pm 0.06 $ & --- & ---\cr
ux40-42 & 15 19 42.15 & 23 53 31.8 & 23.53 & \hbox{No signal} & --- & ---\cr
ux40-44 & 15 19 40.11 & 23 53 06.0 & 24.61 & $ 20.43 \pm 0.43 $ & --- & ---\cr
ux40-45 & 15 19 41.52 & 23 53 12.4 & 22.16 & $ 19.27 \pm 0.15 $ & --- & ---\cr
ux40-46 & 15 19 42.33 & 23 53 03.6 & 99.99 & $ 20.93 \pm 0.65 $ & --- & ---\cr
ux40-47 & 15 19 43.52 & 23 52 29.2 & 22.91 & $ 23.33 \pm 8.21 $ & --- & ---\cr
ux40-64 & 15 19 42.15 & 23 52 12.8 & 24.21 & $ 19.99 \pm 0.47 $ & --- & ---\cr
ux40-65 & 15 19 43.20 & 23 52 17.5 & 22.85 & \hbox{No signal} & --- & ---\cr
ux40-66 & 15 19 43.51 & 23 52 13.3 & 23.21 & $ 19.98 \pm 0.59 $ & --- & ---\cr
uy00-28 & 14 16 18.16 & 11 32 24.6 & 20.42 & $ 18.55 \pm 0.18 $ & 0.41 & 0.04\cr
uy00-30 & 14 16 18.29 & 11 32 19.9 & 20.87 & $ 18.70 \pm 0.21 $ & 0.27 & 0.02\cr
uy00-45 & 14 16 17.59 & 11 31 56.3 & 20.66 & $ 19.04 \pm 0.17 $ & 0.37 & 0.06\cr
uy00-46 & 14 16 17.03 & 11 32 25.5 & 22.57 & $ 20.82 \pm 0.94 $ & --- & ---\cr
uy00-47 & 14 16 17.86 & 11 31 23.4 & 21.87 & \hbox{No signal} & --- & ---\cr
uy00-48 & 14 16 17.68 & 11 31 26.3 & 23.22 & \hbox{No signal} & --- & ---\cr
uy00-49 & 14 16 16.92 & 11 32 08.7 & 22.63 & $ 18.77 \pm 0.11 $ & --- & ---\cr
uy00-50 & 14 16 17.43 & 11 31 25.8 & 22.32 & $ 18.97 \pm 0.16 $ & --- & ---\cr
uy00-51 & 14 16 17.29 & 11 31 36.3 & 23.51 & $ 20.30 \pm 0.48 $ & --- & ---\cr
uy00-52 & 14 16 16.41 & 11 31 57.8 & 21.26 & $ 17.74 \pm 0.04 $ & 0.30 & 0.21\cr
uy00-53 & 14 16 16.33 & 11 31 55.8 & 21.25 & $ 18.06 \pm 0.06 $ & --- & ---\cr
uy00-54 & 14 16 16.66 & 11 31 33.6 & 21.45 & $ 18.87 \pm 0.13 $ & 0.16 & -0.00\cr
uy00-55 & 14 16 15.68 & 11 32 11.3 & 22.43 & $ 20.67 \pm 0.65 $ & --- & ---\cr
uy00-56 & 14 16 16.27 & 11 31 25.1 & 99.99 & \hbox{No signal} & --- & ---\cr
uy00-57 & 14 16 15.51 & 11 32 12.4 & 22.51 & $ 20.37 \pm 0.47 $ & --- & ---\cr
uy00-58 & 14 16 15.92 & 11 31 39.5 & 21.60 & $ 20.44 \pm 0.52 $ & 0.32 & 0.02\cr
uy00-59 & 14 16 15.47 & 11 32 04.7 & 20.04 & $ 17.54 \pm 0.04 $ & 0.49 & 0.07\cr
uy00-60 & 14 16 15.63 & 11 31 55.0 & 22.52 & $ 20.65 \pm 0.65 $ & --- & ---\cr
uy00-61 & 14 16 15.07 & 11 32 19.5 & 20.77 & $ 18.81 \pm 0.14 $ & 0.25 & -0.00\cr
uy00-62 & 14 16 15.79 & 11 31 32.9 & 22.54 & $ 20.07 \pm 0.36 $ & --- & ---\cr
uy00-63 & 14 16 15.35 & 11 31 58.3 & 22.03 & $ 19.57 \pm 0.24 $ & --- & ---\cr
uy00-64 & 14 16 15.73 & 11 31 16.2 & 21.50 & $ 18.79 \pm 0.15 $ & 0.41 & 0.04\cr
uy00-65 & 14 16 15.41 & 11 31 36.1 & 23.57 & \hbox{No signal} & --- & ---\cr
uy00-66 & 14 16 15.11 & 11 31 46.2 & 21.52 & $ 19.10 \pm 0.15 $ & 0.56 & 0.00\cr
uy00-67 & 14 16 14.78 & 11 31 55.5 & 21.33 & $ 18.81 \pm 0.11 $ & 0.48 & 0.05\cr
uy00-68 & 14 16 14.75 & 11 31 49.6 & 23.17 & \hbox{No signal} & --- & ---\cr
uy00-69 & 14 16 15.03 & 11 31 21.1 & 21.85 & $ 19.92 \pm 0.37 $ & 0.28 & 0.03\cr
uy00-70 & 14 16 14.31 & 11 31 55.2 & 21.29 & $ 19.06 \pm 0.14 $ & 0.37 & 0.02\cr
uy00-71 & 14 16 14.24 & 11 31 34.0 & 22.12 & $ 18.70 \pm 0.11 $ & --- & ---\cr
uy00-72 & 14 16 13.39 & 11 32 06.0 & 19.70 & $ 17.48 \pm 0.05 $ & 0.41 & 0.05\cr
uy00-73 & 14 16 13.62 & 11 31 55.6 & 21.63 & $ 20.30 \pm 0.53 $ & 0.42 & 0.01\cr
uy00-74 & 14 16 13.84 & 11 31 35.9 & 20.77 & $ 19.05 \pm 0.17 $ & 0.42 & 0.08\cr
uy00-75 & 14 16 13.94 & 11 31 24.8 & 20.50 & $ 18.98 \pm 0.15 $ & 0.24 & -0.01\cr
uy00-76 & 14 16 13.57 & 11 31 12.6 & 19.71 & $ 17.77 \pm 0.07 $ & 0.29 & 0.22\cr
uy00-77 & 14 16 13.62 & 11 31 14.6 & 19.85 & $ 17.87 \pm 0.08 $ & --- & ---\cr
uy00-78 & 14 16 13.43 & 11 31 19.3 & 22.68 & $ 20.06 \pm 0.59 $ & --- & ---\cr
uy00-79 & 14 16 13.00 & 11 31 46.7 & 22.61 & $ 21.54 \pm 2.03 $ & --- & ---\cr
uy00-80 & 14 16 13.12 & 11 31 31.7 & 99.99 & $ 22.05 \pm 3.22 $ & --- & ---\cr
uzk0-1 & 12 11 11.82 & 39 27 55.3 & 22.80 & \hbox{No signal} & --- & ---\cr
uzk0-2 & 12 11 12.81 & 39 27 10.4 & 22.06 & \hbox{No signal} & --- & ---\cr
uzk0-3 & 12 11 12.66 & 39 27 27.9 & 22.82 & $ 19.11 \pm 0.19 $ & --- & ---\cr
uzk0-4 & 12 11 12.94 & 39 27 36.4 & 21.15 & $ 18.88 \pm 0.16 $ & --- & ---\cr
uzk0-5 & 12 11 12.89 & 39 27 48.8 & 22.18 & $ 19.91 \pm 0.40 $ & --- & ---\cr
uzk0-6 & 12 11 13.73 & 39 27 21.1 & 22.66 & $ 21.44 \pm 1.64 $ & --- & ---\cr
uzk0-7 & 12 11 13.89 & 39 27 15.1 & 22.26 & $ 20.63 \pm 0.68 $ & --- & ---\cr
uzk0-8 & 12 11 14.14 & 39 27 23.4 & 20.29 & $ 17.54 \pm 0.04 $ & --- & ---\cr
uzk0-9 & 12 11 13.87 & 39 27 49.3 & 21.67 & $ 18.60 \pm 0.13 $ & --- & ---\cr
uzk0-10 & 12 11 14.01 & 39 27 52.8 & 19.84 & $ 16.80 \pm 0.02 $ & 0.43 & 0.20\cr
uzk0-11 & 12 11 14.66 & 39 27 38.3 & 22.15 & $ 20.27 \pm 0.60 $ & --- & ---\cr
uzk0-12 & 12 11 14.79 & 39 27 42.1 & 21.69 & $ 18.74 \pm 0.14 $ & 0.33 & -0.10\cr
uzk0-13 & 12 11 14.35 & 39 28 19.7 & 22.87 & $ 20.08 \pm 0.63 $ & --- & ---\cr
uzk0-14 & 12 11 16.18 & 39 27 20.0 & 21.40 & $ 19.04 \pm 0.16 $ & 0.42 & -0.03\cr
uzk0-15 & 12 11 14.95 & 39 28 23.7 & 20.53 & $ 17.90 \pm 0.09 $ & --- & ---\cr
uzk0-16 & 12 11 16.21 & 39 27 29.3 & 23.27 & $ 20.35 \pm 0.58 $ & --- & ---\cr
uzk0-17 & 12 11 16.86 & 39 27 33.6 & 23.08 & \hbox{No signal} & --- & ---\cr
uzk0-18 & 12 11 16.10 & 39 28 13.8 & 21.66 & $ 19.77 \pm 0.40 $ & 0.28 & 0.08\cr
uzk0-19 & 12 11 17.35 & 39 27 24.4 & 22.78 & $ 19.98 \pm 0.48 $ & --- & ---\cr
uzk0-20 & 12 11 16.88 & 39 27 52.7 & 21.41 & $ 18.25 \pm 0.09 $ & 0.50 & 0.03\cr
uzk0-21 & 12 11 16.89 & 39 27 49.9 & 21.90 & $ 19.02 \pm 0.19 $ & --- & ---\cr
uzk0-22 & 12 11 16.28 & 39 28 21.8 & 23.10 & \hbox{No signal} & --- & ---\cr
uzk0-23 & 12 11 17.41 & 39 27 35.6 & 22.95 & $ 20.05 \pm 0.57 $ & --- & ---\cr
uzk0-24 & 12 11 17.64 & 39 27 36.6 & 23.20 & $ 19.94 \pm 0.55 $ & --- & ---\cr
uzk0-25 & 12 11 16.90 & 39 28 12.8 & 22.90 & \hbox{No signal} & --- & ---\cr
uzk0-26 & 12 11 18.01 & 39 27 24.6 & 99.99 & \hbox{No signal} & --- & ---\cr
uzk0-27 & 12 11 17.47 & 39 27 55.8 & 23.64 & $ 19.42 \pm 0.33 $ & --- & ---\cr
uzk0-28 & 12 11 17.13 & 39 28 14.3 & 24.54 & \hbox{No signal} & --- & ---\cr
uzk0-29 & 12 11 17.74 & 39 27 42.3 & 22.93 & \hbox{No signal} & --- & ---\cr
uzk0-30 & 12 11 18.19 & 39 27 27.8 & 99.99 & $ 24.99 \pm 67.84 $ & --- & ---\cr
uzk0-31 & 12 11 16.01 & 39 27 07.3 & 22.93 & $ 19.65 \pm 0.28 $ & --- & ---\cr
uzk0-32 & 12 11 16.44 & 39 27 06.3 & 22.79 & $ 20.27 \pm 0.50 $ & --- & ---\cr
uzk0-33 & 12 11 17.90 & 39 27 07.7 & 22.67 & $ 19.31 \pm 0.27 $ & --- & ---\cr
uzk0-34 & 12 11 15.85 & 39 27 01.1 & 22.48 & $ 19.84 \pm 0.34 $ & --- & ---\cr
uzk0-35 & 12 11 18.54 & 39 27 05.0 & 99.99 & \hbox{No signal} & --- & ---\cr
uzk0-36 & 12 11 14.86 & 39 26 53.5 & 19.98 & $ 17.03 \pm 0.03 $ & 0.50 & 0.05\cr
uzk0-37 & 12 11 15.64 & 39 26 54.9 & 21.17 & $ 18.35 \pm 0.09 $ & 0.36 & 0.05\cr
uzk0-38 & 12 11 14.16 & 39 26 49.2 & 22.52 & $ 20.88 \pm 0.88 $ & --- & ---\cr
uzk0-39 & 12 11 15.07 & 39 26 50.1 & 23.41 & $ 22.46 \pm 3.88 $ & --- & ---\cr
uzk0-40 & 12 11 18.10 & 39 26 54.5 & 24.18 & $ 19.31 \pm 0.22 $ & --- & ---\cr
uzk0-41 & 12 11 15.17 & 39 26 46.2 & 22.12 & $ 20.63 \pm 0.71 $ & --- & ---\cr
uzk0-42 & 12 11 18.39 & 39 26 55.1 & 26.75 & $ 19.00 \pm 0.18 $ & --- & ---\cr
uzk0-43 & 12 11 16.69 & 39 26 48.1 & 21.92 & $ 20.07 \pm 0.42 $ & --- & ---\cr
uzk0-44 & 12 11 17.83 & 39 26 49.0 & 22.56 & $ 19.63 \pm 0.27 $ & --- & ---\cr
uzk0-45 & 12 11 15.29 & 39 26 34.7 & 22.62 & $ 19.96 \pm 0.38 $ & --- & ---\cr
uzk0-46 & 12 11 13.00 & 39 26 27.9 & 99.99 & $ 20.28 \pm 0.47 $ & --- & ---\cr
uzk0-47 & 12 11 17.90 & 39 26 40.8 & 23.21 & $ 21.35 \pm 1.34 $ & --- & ---\cr
uzk0-48 & 12 11 18.74 & 39 26 41.5 & 23.45 & \hbox{No signal} & --- & ---\cr
uzk0-49 & 12 11 17.44 & 39 26 34.6 & 23.47 & $ 19.08 \pm 0.18 $ & --- & ---\cr
uzk0-50 & 12 11 17.37 & 39 26 29.9 & 22.31 & $ 18.32 \pm 0.09 $ & --- & ---\cr
uzk0-51 & 12 11 17.96 & 39 26 25.9 & 22.92 & $ 20.01 \pm 0.38 $ & --- & ---\cr
uzk0-52 & 12 11 19.02 & 39 26 27.6 & 20.53 & $ 18.71 \pm 0.16 $ & 0.40 & 0.07\cr
uzk0-53 & 12 11 18.60 & 39 26 21.7 & 22.63 & $ 21.18 \pm 1.27 $ & --- & ---\cr
uzk0-54 & 12 11 15.83 & 39 26 12.8 & 22.80 & $ 19.94 \pm 0.37 $ & --- & ---\cr
uzk0-55 & 12 11 16.52 & 39 26 14.0 & 24.10 & $ 20.55 \pm 0.65 $ & --- & ---\cr
uzk0-56 & 12 11 18.95 & 39 26 19.0 & 25.68 & $ 20.78 \pm 1.03 $ & --- & ---\cr
uzk0-57 & 12 11 16.96 & 39 26 08.2 & 22.77 & $ 19.09 \pm 0.19 $ & --- & ---\cr
uzk0-58 & 12 11 16.23 & 39 26 04.4 & 22.43 & $ 18.90 \pm 0.19 $ & --- & ---\cr
uzk0-59 & 12 11 17.85 & 39 26 09.1 & 23.37 & $ 19.52 \pm 0.27 $ & --- & ---\cr
uzk0-60 & 12 11 15.95 & 39 26 02.7 & 22.73 & $ 18.60 \pm 0.15 $ & --- & ---\cr
uzk0-61 & 12 11 19.25 & 39 26 09.9 & 26.06 & $ 20.55 \pm 1.04 $ & --- & ---\cr
uzk0-62 & 12 11 12.19 & 39 26 43.4 & 99.99 & \hbox{No signal} & --- & ---\cr
uzk0-63 & 12 11 12.33 & 39 26 35.7 & 24.23 & \hbox{No signal} & --- & ---\cr
uzk0-64 & 12 11 12.53 & 39 26 19.5 & 99.99 & \hbox{No signal} & --- & ---\cr
uzk0-65 & 12 11 12.38 & 39 26 23.1 & 23.92 & $ 20.40 \pm 0.52 $ & --- & ---\cr
uzk0-66 & 12 11 12.22 & 39 26 31.0 & 24.02 & $ 20.05 \pm 0.39 $ & --- & ---\cr
uzk0-67 & 12 11 11.99 & 39 26 38.3 & 99.99 & \hbox{No signal} & --- & ---\cr
uzk0-68 & 12 11 12.15 & 39 26 20.5 & 22.57 & $ 19.84 \pm 0.31 $ & --- & ---\cr
uzk0-69 & 12 11 11.71 & 39 26 42.5 & 24.04 & $ 21.28 \pm 1.15 $ & --- & ---\cr
uzk0-70 & 12 11 12.21 & 39 26 14.1 & 23.52 & $ 21.75 \pm 1.79 $ & --- & ---\cr
uzk0-71 & 12 11 11.79 & 39 26 27.6 & 22.63 & $ 19.81 \pm 0.30 $ & --- & ---\cr
uzk0-72 & 12 11 12.47 & 39 25 52.5 & 24.34 & \hbox{No signal} & --- & ---\cr
uzk0-73 & 12 11 11.38 & 39 26 43.5 & 22.88 & $ 20.70 \pm 0.67 $ & --- & ---\cr
uzk0-74 & 12 11 12.19 & 39 25 51.7 & 22.95 & $ 20.00 \pm 0.43 $ & --- & ---\cr
uzk0-75 & 12 11 11.92 & 39 25 59.3 & 21.96 & $ 19.72 \pm 0.29 $ & 0.30 & 0.03\cr
uzk0-76 & 12 11 11.76 & 39 26 08.0 & 22.56 & $ 19.94 \pm 0.32 $ & --- & ---\cr
uzk0-77 & 12 11 11.58 & 39 25 51.6 & 22.91 & $ 19.34 \pm 0.21 $ & --- & ---\cr
uzk0-78 & 12 11 11.39 & 39 25 48.9 & 99.99 & $ 18.58 \pm 0.11 $ & --- & ---\cr
uzk0-79 & 12 11 10.33 & 39 26 54.7 & 23.94 & $ 22.46 \pm 4.71 $ & --- & ---\cr
uzk0-80 & 12 11 10.73 & 39 26 23.0 & 20.52 & $ 17.50 \pm 0.04 $ & 0.53 & 0.04\cr
uzk0-81 & 12 11 10.42 & 39 26 34.1 & 20.52 & $ 17.64 \pm 0.04 $ & 0.41 & 0.04\cr
uzk0-82 & 12 11 10.16 & 39 26 47.7 & 21.33 & $ 18.05 \pm 0.07 $ & 0.39 & 0.05\cr
uzk0-83 & 12 11 11.15 & 39 25 57.9 & 22.58 & $ 22.57 \pm 3.60 $ & --- & ---\cr
uzk0-84 & 12 11 09.52 & 39 26 33.9 & 21.92 & $ 18.80 \pm 0.12 $ & 0.23 & 0.30\cr
uzk0-85 & 12 11 10.30 & 39 25 48.4 & 22.72 & $ 21.30 \pm 1.35 $ & --- & ---\cr
uzk0-86 & 12 11 09.40 & 39 26 11.5 & 22.90 & $ 19.84 \pm 0.29 $ & --- & ---\cr
uzk0-87 & 12 11 08.72 & 39 26 34.6 & 23.00 & $ 21.55 \pm 1.47 $ & --- & ---\cr
uzk0-88 & 12 11 08.86 & 39 26 26.3 & 22.00 & $ 19.83 \pm 0.31 $ & --- & ---\cr
uzk0-89 & 12 11 08.85 & 39 26 23.8 & 22.18 & $ 19.50 \pm 0.22 $ & --- & ---\cr
uzk0-90 & 12 11 08.87 & 39 25 47.3 & 21.41 & $ 18.68 \pm 0.12 $ & 0.52 & 0.19\cr
uzk0-91 & 12 11 08.57 & 39 26 06.0 & 22.32 & $ 19.37 \pm 0.19 $ & --- & ---\cr
uzk0-92 & 12 11 08.62 & 39 26 03.5 & 23.06 & $ 19.90 \pm 0.31 $ & --- & ---\cr
uzk0-93 & 12 11 07.98 & 39 26 27.9 & 22.57 & $ 18.74 \pm 0.11 $ & --- & ---\cr
uzk0-94 & 12 11 08.33 & 39 26 06.5 & 22.50 & $ 20.92 \pm 0.81 $ & --- & ---\cr
uzk0-95 & 12 11 07.65 & 39 26 26.7 & 22.55 & $ 20.03 \pm 0.39 $ & --- & ---\cr
} 
%\end{document} 

}

\vfill\eject
\onecolumn
\bigskip\bigskip
{\bf TABLE 3.}\quad Statstical comparison of $I-K$ distribution with models.
\bigskip\bigskip
\def\sp{\hspace*{0.2em}}
\def\med{$Med$}
\def\dsh{\hbox{---}}

\begin{tabular}{lcccccc}
\large Model & \multicolumn{3}{c}{\large All} &  \multicolumn{3}{c}{\large E/S0}  \\
      & \sp\PKS & \XS & \med \sp&\sp \PKS & \XS& \med \\
No-Evolution SEDs $\Omega=0$ &  -12.67 &  1.60 &  2.85 & -6.95 &  2.23 &  3.20 \\
No-Evolution SEDs $\Omega=1$ &  -12.30 &  1.99 &  2.85 & -7.09 &  3.08 &  3.25 \\
PBZ Luminosity Evolution   $\Omega=0$ & -2.88 &  1.36 &  2.60 & -6.52 &  1.34 &  3.25  \\
PBZ Luminosity Evolution   $\Omega=1$ & -1.54 &  1.52 &  2.50 & -6.94 &  1.32 &  3.35  \\
\noalign{\vspace{0.5cm}}
\large Model & \multicolumn{3}{c}{\large Spiral} & \multicolumn{3}{c}{\large Irr/Pec} \\
      & \sp\PKS & \XS & \med \sp&\sp \PKS & \XS& \med \\
No-Evolution SEDs $\Omega=0$ &   -6.45 &  0.90 &  2.80 & -4.81 &  5.87 &  2.45\\
No-Evolution SEDs $\Omega=1$ &   -6.37 &  1.10 &  2.80 & -4.45 &  6.96 &  2.45\\
PBZ Luminosity Evolution   $\Omega=0$ & -3.27 &  0.88 &  2.50 & -8.65 &  4.63 &  2.10 \\
PBZ Luminosity Evolution   $\Omega=1$ & -5.69 &  1.04 &  2.40 & -8.68 &  5.22 &  2.05 \\
`Eternally Middle-Aged' ($\Omega=1$)  & \dsh  & \dsh  & \dsh  & -2.73 &  1.83 &  2.55 \\
\end{tabular}

\label{lastpage}

\end{document}